\documentclass[pra,onecolumn,preprint,11pt,showpacs,citeautoscript,footinbib,eqsecnum]{revtex4-2}
\usepackage[T1]{fontenc}
\usepackage[utf8]{inputenc}
\usepackage{lipsum}
\usepackage{amsmath,bm}
\usepackage{amssymb}
\usepackage{bbm}
\usepackage{braket}
\usepackage{graphicx}
\usepackage{xcolor}
\usepackage{pifont}
\usepackage[mathscr]{euscript}
\usepackage[shortlabels]{enumitem}
\usepackage[justification=justified]{subcaption}
\usepackage{float}
\usepackage{dsfont}
\usepackage{comment}
\usepackage{subcaption}
\usepackage{slashed}

\usepackage[papersize={8.5in,11in}]{geometry}
\usepackage{tikz}
\usetikzlibrary{decorations.markings}
\usetikzlibrary{decorations.pathmorphing}
\tikzset{snake it/.style={decorate, decoration=snake}}
\tikzset{->-/.style={decoration={
  markings,
  mark=at position #1 with {\arrow{>}}},postaction={decorate}}}
\tikzset{-<-/.style={decoration={
  markings,
  mark=at position #1 with {\arrow{<}}},postaction={decorate}}}

\usepackage{ragged2e}
\DeclareCaptionJustification{justified}{\justifying}
\allowdisplaybreaks

\usepackage[colorlinks=true]{hyperref}  
\hypersetup{
    unicode=false,          
    pdftoolbar=true,        
    pdfmenubar=true,        
    pdffitwindow=false,     
    pdfstartview={FitH},    
    pdftitle={Flstar},    
    pdfauthor={Christos, Sachdev, Luo},     
    pdfsubject={},   
    pdfcreator={},   
    pdfproducer={}, 
    pdfkeywords={} {} {}, 
    pdfnewwindow=true,      
    colorlinks=true,       
    linkcolor=blue, 
    citecolor=blue,        
    filecolor=magenta,      
    urlcolor=blue           
}

\newcommand{\be}{\begin{equation}}
\newcommand{\ee}{\end{equation}}
\newcommand{\beq}{\begin{equation}}
\newcommand{\eeq}{\end{equation}}
\renewcommand{\i}{\mathrm{i}}
\newcommand{\tr}{\text{Tr}}
\renewcommand{\L}{\mathcal{L}}

\newcommand{\pdagger}{{\phantom{\dagger}}}

\def\bea{\begin{eqnarray}}
\def\eea{\end{eqnarray}}

\newcommand{\vi}{{\boldsymbol{i}}}
\newcommand{\vj}{{\boldsymbol{j}}}
\newcommand{\vk}{{\boldsymbol{k}}}
\newcommand{\vl}{{\boldsymbol{l}}}

\newcommand{\p}{\partial}

\begin{document}
\preprint{\href{https://arxiv.org/abs/2402.09502}{arXiv:2402.09502}}

\title{Deconfined quantum criticality of \\nodal $d$-wave superconductivity, N\'eel order, and charge order\\ on the square lattice at half-filling}

\begin{abstract} 
We consider a SU(2) lattice gauge theory on the square lattice, with a single fundamental complex fermion and a single fundamental complex boson on each lattice site. Projective symmetries of the gauge-charged fermions are chosen so that they match with those of the spinons of the $\pi$-flux spin liquid. Global symmetries of all gauge-invariant observables are chosen to match with those of the particle-hole symmetric electronic Hubbard model at half-filling. Consequently, both the fundamental fermion and fundamental boson move in an average background $\pi$-flux, their gauge-invariant composite is the physical electron, and eliminating gauge fields in a strong gauge-coupling expansion yields an effective extended Hubbard model for the electrons. The SU(2) gauge theory displays several confining/Higgs phases: a nodal $d$-wave superconductor, and states with N\'eel, valence-bond solid, charge, or staggered current orders. There are also a number of quantum phase transitions between these phases which are very likely described by 2+1 dimensional deconfined conformal gauge theories, and we present large flavor expansions for such theories. These include the phenomenologically attractive case of a transition between a conventional insulator with a charge gap and N\'eel order, and a conventional $d$-wave superconductor with gapless Bogoliubov quasiparticles at 4 nodal points in the Brillouin zone. We also apply our approach to the honeycomb lattice, where we find a bicritical point at the junction of N\'eel, valence bond solid (Kekul\'e), and Dirac semi-metal phases.
\end{abstract}

\author{Maine Christos}
\affiliation{Department of Physics, Harvard University, Cambridge MA 02138, USA}

\author{Henry Shackleton}
\affiliation{Department of Physics, Harvard University, Cambridge MA 02138, USA}

\author{Subir Sachdev}
\affiliation{Department of Physics, Harvard University, Cambridge MA 02138, USA}

\author{Zhu-Xi Luo}
\affiliation{Department of Physics, Harvard University, Cambridge MA 02138, USA}

\maketitle
\newpage
\linespread{1.05}
\tableofcontents

\section{Introduction}

The cuprate high temperature superconductors display a complex phase diagram involving low temperature ($T$) phases with $d$-wave superconductivity, N\'eel antiferromagnetic order, and charge order, and the higher $T$ pseudogap and strange metals \cite{keimernature}. The remarkable pseudogap metal phase is of central importance, and many of its properties can be described by a model of hole pocket Fermi surfaces \cite{SS94,WenLee96,WenLee98,Stanescu_2006,Berthod_2006,Yang_2006,Kaul07,KaulKim07,Sakai_2009,Qi_2010,Mei_2012,Robinson_2019,Ancilla1,Skolimowski_2022,Fabrizio23,Giorgio23,PNAS_pseudo,Lucila23}. Such Fermi surfaces enclose an area distinct from the Luttinger volume, and this requires the presence of a background spin liquid, realizing a state that has been called a `fractionalized Fermi liquid' (FL*) \cite{FLS1,FLS2,Arun04}.
Recent works \cite{PNAS_pseudo,Christos23} have proposed that the low $T$ cuprate  phase diagram can be understood from a theory of the confining instabilities of a FL* state with a `$\pi$-flux' critical spin liquid on the square lattice. The critical spin liquid emerges from a background into a central role in such confining transitions, and a detailed understanding of its role then becomes a central ingredient in unraveling the mysteries of the cuprate phase diagram.

An important feature of the FL* theory is that its fractionalized excitations have the same basic structure as that in a Mott insulator at half-filling, even though the pseudogap state is at non-zero doping. The doping is accounted for by the hole pocket Fermi surfaces, which are coupled to the spin liquid. Given this relatively innocuous influence of non-zero doping, the present paper will investigate a simpler model which remains at half-filling, but has the same set of conventional symmetry-breaking phases without fractionalization at low temperatures, as at non-zero doping: a $d$-wave superconductor with 4 nodal points for Bogoliubov quasiparticles, and conventional states with N\'eel, valence-bond solid, charge, or staggered current orders.
There are quantum phase transitions between these states which are very likely described by deconfined critical points, allowing a systematic study of associated critical spin liquids. 
Our simpler model should be amenable to numerical simulations by the well-developed methods of lattice gauge theory of relativistic systems \cite{LGT_book}, and shed light on the role of spin liquids in the phase diagram of the cuprates.

We begin by noting a few recent developments which relate to the FL*-confinement proposal of Ref.~\onlinecite{PNAS_pseudo}:
\begin{itemize}
\item 
Angle-dependent magnetoresistance measurements on the underdoped cuprates \cite{Fang2020} are consistent with hole pocket Fermi surfaces \cite{SS94,WenLee96,WenLee98,Stanescu_2006,Berthod_2006,Yang_2006,Kaul07,KaulKim07,Sakai_2009,Qi_2010,Mei_2012,Robinson_2019,Ancilla1,Skolimowski_2022,Fabrizio23,Giorgio23,PNAS_pseudo,Lucila23}.
\item 
A long-standing issue with the hole pocket model of the pseudogap metal  is that the pairing of quasiparticles around the hole pocket leads to a $d$-wave superconductors with {\it eight\/} nodal points \cite{Moon11}. This problem can be resolved by {\it not\/} viewing the onset of superconductivity from the pseudogap normal state as a BCS-like pairing of electronic quasiparticles on Fermi surface. 
Instead, the spin liquid of the pseudogap already features a singlet pairing of electrons \cite{PWA87}, and we 
should consider the onset of superconductivity as a confining transition of the $\pi$-flux spin liquid by the condensation of a fundamental Higgs scalar. (In both viewpoints, the non-zero temperature transition of the onset of superconductivity remains in the Kosterlitz-Thouless universality class.) Then the fermionic spinon nodal points of the spin liquid annihilate four of the nodal points descending from the hole pockets, and we obtain a $d$-wave superconductor with {\it four\/} nodal points \cite{ChatterjeeSS16,Christos23}, as is expected in a conventional BCS state. Moreover, the large velocity anisotropy of the nodal quasiparticles is easily obtained in this approach. 
\item Recent work \cite{BCS} has proposed an explanation of the high-field quantum oscillations by considering the influence of the fermionic spinons across the transition from FL* to the field-induced charge-ordered state. 
\item
Photoemission observations in the electron-doped cuprates \cite{Shen_2023} show a gap maximum at an intermediate wavevector away from the edge of the Brillouin zone, and not on the Fermi surface. This feature is also obtained as a consequence of the background spin liquid \cite{Christos23}. Indeed, even when the pseudogap metal has no Fermi surfaces intersecting the zone diagonals, the resulting $d$-wave superconductor still has 4 nodal points along the zone diagonals, and these are directly descended from the nodal spinons of the underlying spin liquid \cite{Christos23}.
\item 
Numerical fuzzy sphere and other studies have found evidence for $\pi$-flux spin liquid criticality, which ultimately gives way either to `pseudo-criticality' \cite{Zhou:2023qfi} or nearby multi-criticality \cite{Zhao20,Chester:2023njo,Meng23,Sandvik24}. In contrast, the commonly used `staggered flux' spin liquid \cite{LeeRMP} is expected to be strongly unstable to a trivial monopole \cite{Alicea08,Song:2018ial}.
\item 
Numerical studies \cite{Sandvik18,Becca20,Imada20,Gu22,Qian:2023fkv} of $S=1/2$ square lattice antiferromagnets with first- and second-neighbor exchange interactions (the $J_1$-$J_2$ antiferromagnet) display a transition from the N\'eel state to valence bond solid order \cite{NRSS89,NRSS91}, across an intermediate spin-liquid regime which is likely described by the $\pi$-flux spin liquid \cite{DQCP3}. 
A gapless $\mathbb{Z}_2$ spin liquid has also been proposed for this intermediate regime, and this can be obtained naturally by condensing Higgs fields on the $\pi$-flux spin liquid \cite{YingRanThesis,RanWen06,Shackleton:2021fdh,Shackleton:2022zzm} (the model studied in the present paper can be easily extended to include these Higgs fields, but we will not present the extension here \footnote{The theory of Ref.~\onlinecite{Shackleton:2021fdh} can exhibit a bicritical point where the N\'eel, VBS, and gapless $\mathbb{Z}_2$ spin liquids meet. This bicritical point can realize SO(5)-symmetric bicritical point in the studies of Ref.~\onlinecite{Chester:2023njo,Sandvik24} if the Yukawa couplings between the Higgs fields and spinons are irrelevant at the bicritical point.}).
Doping this square lattice spin liquid has recently been shown \cite{Jiang21,Jiang23} to lead to robust $d$-wave superconductivity, and this establishes a close connection between the $\pi$-flux phase and $d$-wave superconductivity \cite{ZhangRice-dSC,Senthil_Ivanov}. 
\item 
Nuclear magnetic resonance experiments on YBa$_2$Cu$_3$O$_y$ \cite{Julien_2024} show the appearance of a secondary spin gap which is possibly connected to the appearance of charge order. This can be associated with the gapping out of the spinon excitations upon a confining transition to charge order, as we study in a simplified model in this paper.
\item 
Magnetotransport studies in HgBa$_2$Ca$_2$Cu$_3$O$_{8+ \delta}$ \cite{Proust_2024} indicate a direct transitions between magnetic and charge ordered states. Such direct transitions are possible across deconfined critical points considered here. 
\end{itemize}

The `$\pi$-flux' critical spin liquid is described by a theory of fermionic spinons with $N_f =2$ massless Dirac points in their dispersion coupled to a SU(2) gauge field \cite{Affleck1988}. This state also has a dual description \cite{DQCP3} in terms of the critical $\mathbb{CP}^1$ theory of the bosonic spinons \cite{NRSS89}.
These dual descriptions are important in understanding the low temperature states of the cuprate phase diagram as confinement/Higgs transitions of this spin liquid:
\begin{itemize}
\item[({\it i\/})] The onset of N\'eel order is described by the Higgs condensate of the bosonic spinons in the $\mathbb{CP}^1$ theory \cite{Nikolaenko23}, or equivalently, by the confinement of the SU(2) gauge field of the fermionic spinon theory.
\item[({\it ii\/})] The onset of $d$-wave superconductivity with nodal Bogoliubov quasiparticles \cite{ChatterjeeSS16}, along with the onset of charge order, is described by the Higgs condensation of a charge $e$, SU(2) fundamental boson $B$ (introduced in Refs.~\onlinecite{WenLee96,LeeRMP}) of the fermionic spinon theory.
\end{itemize}

As noted above, this paper will study a simpler limit of the theory of Ref.~\onlinecite{PNAS_pseudo}. We will move from the system at non-zero doping, and instead consider only the half-filled square lattice with a particle-hole symmetric Hamiltonian. Rather than introducing superconductivity and charge-order by doping, we will explore the onset of such phases at half-filling as may be induced by reducing the 
Hubbard $U$ \cite{Raghu10}, or by introducing additional short-range interactions including pair-hopping terms \cite{AssaadImada,Xu:2020qbj}.

\begin{figure}
    \centering
    \includegraphics[width=6.3in]{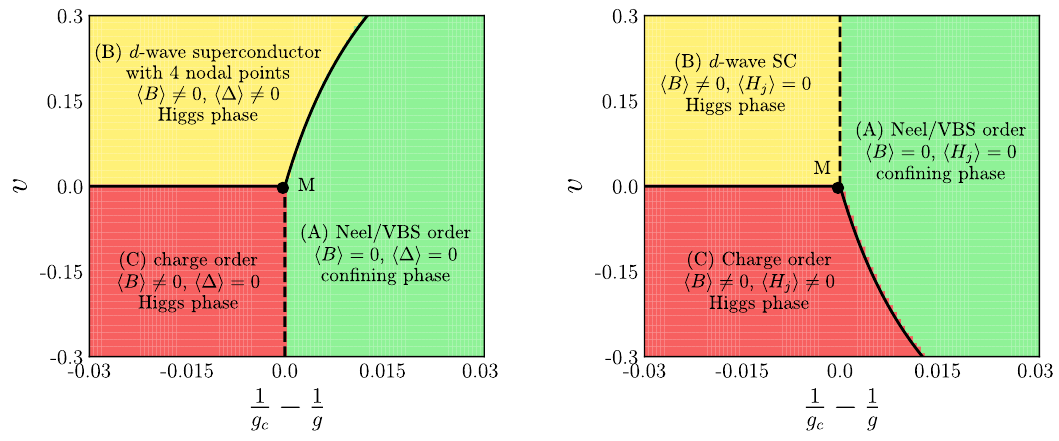}
    \caption{ We are interested in a SU(2) gauge theory with $N_f$ fundamental Dirac fermions, and $N_b=2$ fundamental complex scalars. We show phase diagrams of two distinct large $N_f$ and $N_b$ limits, with $N_f/N_b$ fixed. First order phase transitions are denoted with a solid line while second order phase transitions are denoted with a dashed line. (a) Phase diagram of the theory $\mathcal{L}_\psi + \mathcal{L}_B$ in (\ref{LF}) and (\ref{e2}). There is a USp$(2N_f) \times$USp$(N_b)\times$U(1) global symmetry for $v\neq 0$.  (b) Phase diagram in an alternative large $N_b$ limit discussed in Appendix~\ref{app:alternative} of the theory $\mathcal{L}_\psi + \widetilde{\mathcal{L}}_B$ in (\ref{LF}) and (\ref{e2aa}), with a USp$(2N_f) \times$SU$(N_b)\times$U(1) global symmetry for $v\neq 0$. The theories in (a) and (b) co-incide along the line $v=0$, when they both have USp$(2N_f) \times$USp$(2 N_b)$ global symmetry. The two theories are also identical for the physically interesting case with $N_f = N_b = 2$ for all $v$.}
    \label{fig:1}
\end{figure}

At half-filling, there are no hole pocket Fermi surfaces, and this simplifies the treatment of charge fluctuations. The particle-hole symmetry leads to a Lorentz-invariant form for the dispersion of the excitations at low energies.  We will study zero temperature quantum phase transitions between (A) the insulating N\'eel state, (B) a $d$-wave superconductor with 4 gapless nodal quasiparticles, and (C) a state with charge order; see Fig.~\ref{fig:1} for the phase diagrams of the continuum field theories to be introduced in Section~\ref{sec:introqft} and Appendix~\ref{app:alternative}.
This field theory is a SU(2) gauge theory $N_b = 2$ relativistic scalars in addition to the $N_f=2$ massless Dirac fermions of the $\pi$-flux state.

We note an earlier work \cite{RanVishwanath-easyplane} which considered a continuous N\'eel/$d$-wave superconductor quantum transition, but without gapless nodal quasiparticles in the $d$-wave superconductors, and only easy-plane N\'eel order. Also, SU(2) gauge theories of the cuprates have been studied extensively earlier, as reviewed in Ref.~\onlinecite{LeeRMP}, but in reference to a staggered-flux spin liquid which breaks the gauge symmetry to U(1)---we will not consider this spin liquid because it is expected to be unstable to a trivial monopole \cite{Alicea08,Song:2018ial}.

\begin{figure}
    \centering
    \includegraphics[width=4in]{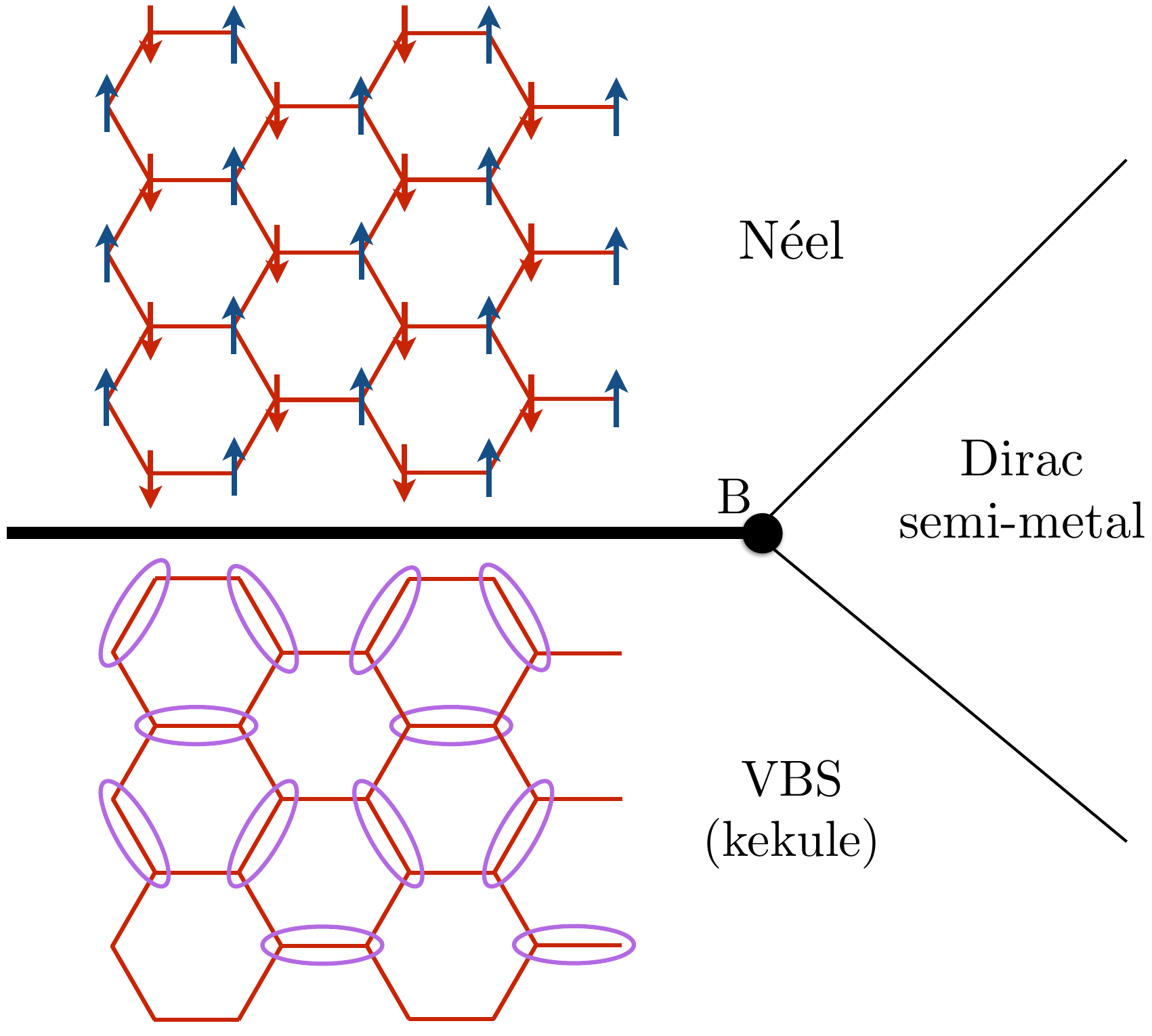}
    \caption{Schematic phase diagram for the SU(2) gauge theory of an extended Hubbard model on the honeycomb lattice. The bicritical point B \cite{KosterlitzNelson} is described by the $N_f=2$, $N_b=1$ SU(2) gauge field theory. The thick line indicates a first-order transition. The thin lines indicate second-order transitions out of the Dirac semi-metal phase which are presumed to be described by Gross-Neveu-Yukawa field theories \cite{Boyack:2020xpe} without gauge fields.}
    \label{fig:honeycomb}
\end{figure}

In Section~\ref{sec:honeycomb}, we will consider the consequences of adding charge fluctuations to the N\'eel-VBS transition on the honeycomb lattice \cite{NRSS90,Fu11} (VBS order is also known as Kekul\'e order on the honeycomb lattice). Following the same procedure as for the square lattice, we find only a Dirac semi-metal phase with no broken symmetry, in contrast to the superconducting and charge-ordered  phases on the square lattice. As shown in Fig.~\ref{fig:honeycomb}, the N\'eel, VBS, and Dirac semi-metal phases of the honeycomb lattice are proposed to meet at a multicritical point, as in the numerical study of the Hubbard model on the honeycomb lattice in Ref.~\onlinecite{Yao19}. In our theory, the multicritical point is bicritical \cite{KosterlitzNelson}, and is
described by the $N_f=2$, $N_b = 1$ case of the SU(2) gauge field theory considered in the body of the paper.
The same field theory was considered earlier by Hermele \cite{Hermele07} for a different proposed transition on the honeycomb lattice. 

Our main results here are obtained by two different large flavor expansions of our SU(2) gauge theory. The resulting phase diagrams in Fig.~\ref{fig:1} contains first-order boundaries, a multi-critical point M where all three phases meet, and second-order transitions between N\'eel/VBS order, charge order, and nodal $d$-wave superconductivity. The multi-critical point M and the second-order transition are described by deconfined critical SU(2) gauge theories. We will determine the scaling dimensions of gauge-invariant N\'eel, valence bond solid (VBS), $d$-wave superconductor, and charge order parameters in these critical theories. 

Of particular interest is the scaling dimension of the gauge-invariant electron operator, which we also determine. This controls the manner in which gapless nodal quasiparticles emerge in the $d$-wave superconductor across the transition from an insulator with a non-zero gap to charged excitations. We summarize the results on scaling dimensions in Table~\ref{tab:dimensions}. 
Ref.~\onlinecite{Christos23} considered a mean-field theory of the corresponding transition in the electron-doped cuprates: in this case, the transition is to a pseudogap-metal, but the nodal region of the Brillouin zone can be gapped in the electron-doped pseudogap metal. Thus our theory has a remarkable feature not present in BCS theory: gapless nodal quasiparticles appear in a superconductor at a momentum which is gapped in the normal state.
As we noted above, Ref.~\onlinecite{Christos23} pointed out connections of this feature to recent photoemission experiments in the electron-doped cuprates \cite{Shen_2023}. 

Section~\ref{sec:lgt} introduces the square lattice SU(2) gauge theory of interest in this paper. This theory is defined in terms of fermionic spinons $f_{\vi \alpha}$, $\alpha = \uparrow, \downarrow$ and charge $e$ bosons $B_{\vi}$ on the sites $\vi$ of the square lattice. Both the fermionic and bosonic matter fields transform as SU(2) gauge fundamentals, and there is also a dynamical SU(2) gauge field on the links of the lattice. We then consider the most general lattice gauge theory for these matter and gauge fields consistent with the projective symmetry transformations of the $\pi$-flux spin liquid, and with gauge-invariant observables having the same symmetry signatures as the Hubbard model with particle-hole symmetry. In the limit of strong gauge couplings, we can perform a strong-coupling expansion of our lattice gauge theory by integrating out the lattice gauge fields \cite{LGT_book}, and this will lead to the extended Hubbard model corresponding to our SU(2) lattice gauge theory. See Chapter 14 of Ref.~\onlinecite{QPMbook} for a simpler example of a conventional theory of gauge-invariant degrees of freedom obtained from a lattice gauge theory of partons. 

Note that our method is the converse of that usually followed in the condensed matter literature. We do not start from a lattice model of correlated electrons, and then obtain a gauge theory by fractionalizing the electrons. Instead, we start from a lattice gauge theory and match it to the electronic problem of interest by general arguments based on gauge invariance and global symmetry. This is a powerful method of incorporating non-perturbative knowledge of a fractionalized state (in our case, the $\pi$-flux spin liquid) in a very general setting.

Section~\ref{sec:introqft} describes the continuum limit of the square lattice gauge theory of Section~\ref{sec:lgt} along the lines of Ref.~\onlinecite{PNAS_pseudo}. This leads to a quantum field theory of $N_f=2$ Dirac fermions and $N_b=2$ complex scalars, both transforming as SU(2) gauge fundamentals. We also discuss the generalizations of this theory to general $N_{f,b}$, and the operators corresponding to the gauge-invariant observables of the Hubbard model.

Section~\ref{sec:spin-gap} examines the nature of fermion-boson couplings in the continuum field theory without any spatial and temporal gradients. We find that there are no allowed terms which are relevant in the large $N_{f,b}$ expansion of critical theories. However, we do need to consider the higher-order formally irrelevant terms because they are important in determining the fate of the spin gap in the Higgs phases where the bosons are condensed.  

Section~\ref{sec:saddlepoint} describes the $N_b = \infty$ saddle points of the continuum theories which lead to the phase diagrams in Fig.~\ref{fig:1}.

Section~\ref{sec:1N} computes the $1/N_{f,b}$ corrections to the scaling dimensions of the $d$-wave superconducting, N\'eel, and charge order parameters, and the electron operator at momenta $(\pm\pi/2, \pm \pi/2)$. This is carried out by the SU(2) gauge theory analog of the computations in Ref.~\onlinecite{KaulSS08} for U(1) gauge theories. 

Section~\ref{sec:honeycomb} describes the extension of our results to the honeycomb lattice.

\section{SU(2) square lattice gauge theory}
\label{sec:lgt}

We begin by recalling the SU(2) square lattice gauge theory of Ref.~\onlinecite{PNAS_pseudo} in the simpler setting of a half-filled square lattice, with no Fermi surfaces in any of the states studied. We also assume a particle-hole symmetry. This lattice gauge theory is likely free of a sign problem in quantum Monte Carlo. 

We write the electron spin operators as
\beq
{C}_{\vi} = \left( \begin{array}{c} {c}_{\vi\uparrow} \\ {c}_{\vi\downarrow}^\dagger \end{array} \right)\,,
\label{eq:Nambu2}
\eeq
on sites $\vi$ of a square lattice. 
We fractionalize the electrons into fermionic spinons $f_{\vi \alpha}$, $\alpha = \uparrow, \downarrow$ and charge $e$ bosons $B_{\vi}$ via \cite{WenLee96}
\beq
C_{\vi} = \mathcal{B}_{\vi}^\dagger \psi_\vi \,,
\label{CBpsi}
\eeq
where
\beq
\psi_{\vi} \equiv \left( \begin{array}{c} f_{\vi\uparrow} \\ f_{\vi\downarrow}^\dagger \end{array} \right)\,,
\label{eq:Nambu1}
\eeq
and 
\beq
B_\vi \equiv \left( \begin{array}{c} B_{1\vi} \\ B_{2\vi} \end{array} \right) \quad, \quad \mathcal{B}_\vi \equiv \left( \begin{array}{cc} B_{1\vi} & - B_{2\vi}^\ast \\ B_{2\vi} & B_{1\vi}^\ast \end{array} \right) \label{Bdef}
\,.
\eeq
This fractionalization introduces a SU(2) gauge symmetry, where 
\beq
\psi_\vi \rightarrow U_\vi \psi_\vi \quad, \quad B_{\vi} \rightarrow U_\vi B_{\vi}\,, \label{gaugedef}
\eeq
under a SU(2) gauge transformation $U_\vi$.

Remarkably, essentially all of the physics of the $\pi$-flux spin liquid phase, and its descendants, studied here are consequences of the SU(2) gauge symmetry, the spin rotation symmetry, and the action of other symmetries on the spinons as summarized in Table.~\ref{PSG1}.
\begin{table}
    \centering
    \begin{tabular}{|c|c|c|c|}
\hline
Symmetry & $c_\alpha$ &$f_{\alpha}$ & $B_a $  \\
\hline 
\hline
$T_x$ &$c_\alpha$ &$(-1)^{y} f_{ \alpha}$ & $(-1)^{y} B_a $ \\
$T_y$ & $c_\alpha$ & $ f_{ \alpha}$ & $ B_a $ \\
$P_x$ & $c_\alpha$ & $(-1)^{x} f_{ \alpha}$  & $(-1)^{x} B_a $ \\
$P_y$ & $c_\alpha$ & $(-1)^{y} f_{ \alpha}$  & $(-1)^{y} B_a $ \\
$P_{xy}$ & $c_\alpha$ &$(-1)^{xy} f_{ \alpha}$ & $(-1)^{x y} B_a $ \\
$\mathcal{T}$ & $\varepsilon_{\alpha\beta} c_\beta$ & $(-1)^{x+y} \varepsilon_{\alpha\beta} f_{ \beta}$  & $(-1)^{x + y} B_a $\\
$\mathcal{C}$ & $(-1)^{x+y} \varepsilon_{\alpha\beta} c_{ \beta}^\dagger$ & $ \varepsilon_{\alpha\beta} f_{ \beta}^\dagger$ & $(-1)^{x+y} B_a^\ast$ \\
\hline
\end{tabular}
    \caption{Projective transformations of the $f_{\vi \alpha}$ spinons and $B_{\vi}$ chargons on lattice sites $\vi = (x,y)$ 
    under the symmetries $T_x: (x,y) \rightarrow (x + 1, y)$; $T_y: (x,y) \rightarrow (x,y + 1)$; 
    $P_x: (x,y) \rightarrow (-x, y)$; $P_y: (x,y) \rightarrow (x, -y)$; $P_{xy}: (x,y) \rightarrow (y, x)$; time-reversal $\mathcal{T}$, and particle-hole symmetry $\mathcal{C}$.
    The indices $\alpha,\beta$ refer to global SU(2) spin, while the index $a=1,2$ refers to gauge SU(2). Also shown are the (non-projective) transformations of the gauge-invariant electron $c_\alpha$.
    }
    \label{PSG1}
\end{table}
The action of the latter symmetries on the $B$ chargons follows from the decomposition (\ref{CBpsi}), and these are also shown in Table~\ref{PSG1}. A key property of Table~\ref{PSG1} is the relation
\beq
T_x T_y = - T_y T_x\,, \label{txty}
\eeq
which ensures $\pi$-flux on both spinons and chargons, and at least two degenerate minima in the dispersion the chargons.

The degrees of freedom of our square lattice gauge theory are one SU(2) fundamental fermion $\psi_\vi$ on each lattice site, one SU(2) fundamental boson $B_\vi$ on each lattice site, and a SU(2) link field $U_{\vi \vj}$ on each nearest-neighbor link of the square lattice. 
We now describe the various terms in the Hamiltonian coupling these degrees of freedom.

The simplest fermion spinon imaginary time ($\tau$) Lagrangian compatible with Table~\ref{PSG1} is
\beq
\mathcal{L} (\psi) = \sum_\vi \psi_\vi^\dagger D_\tau \psi_\vi^\pdagger - i J \sum_{ \langle\vi\vj\rangle  } \left[   \psi_{\vi}^\dagger e_{\vi\vj }^{\vphantom\dagger} U_{\vi \vj}^{\vphantom\dagger} \psi_{\vj }^{\vphantom\dagger} + \vi \leftrightarrow \vj \right]\,, \label{Hs}
\eeq
where $D_\tau$  is a co-variant time derivative, $\vi$,$\vj$ are nearest-neighbors, $J$ is a real coupling constant of order the antiferromagnetic exchange, 
\beq
e_{\vj\vi} = - e_{\vi\vj} 
\eeq 
is a fixed element of the $\mathbb{Z}_2$ center of the gauge SU(2) which ensures $\pi$ flux per plaquette; we choose
\beq
 e_{\vi,\vi+\hat{{\bm x}}}  =  1 \,,\quad
 e_{\vi,\vi+\hat{{\bm y}}}  =  (-1)^{x}      \,, \label{su2ansatz}
\eeq
where $\vi = (x,y)$, $\hat{\bm x} = (1,0)$, $\hat{\bm y} = (0,1)$.
The link field $U_{\vi \vj} = U_{\vj\vi}^\dagger$ is the fluctuating SU(2) lattice gauge field, and the mean-field saddle point of the $\pi$-flux phase is obtained by setting $U_{\vi \vj} = 1$. 
The hopping term in $\mathcal{L}(\psi)$ has been chosen pure imaginary as that ensures a simple coupling to the SU(2) gauge field, along with SU(2) spin rotation invariance. The spin operator on each site ${\bm S}_{\vi} = (1/2) f_{\vi \alpha}^\dagger {\bm \sigma}_{\alpha\beta} f_{\vi \beta}^{\vphantom\dagger}$ (${\bm \sigma}$ are the Pauli matrices) can be expressed in terms of the $\psi_\vi$ in the following SU(2) gauge-invariant combinations:
\beq
2 S_{z\vi} = \psi_\vi^\dagger \psi_\vi^\pdagger -1 \quad, \quad S_{x\vi} - \i S_{y \vi} = -\varepsilon_{ab} \psi_{a \vi} \psi_{b \vi}\,,
\label{SL}
\eeq
where $a,b=1,2$ are SU(2) gauge indices, and $\varepsilon_{ab}$ is unit antisymmetric tensor.
The nearest-neighbor bond energy operator can be identified with each individual term in $\mathcal{L} (\psi)$
\begin{align}
&\mbox{bond energy:~} \left\langle {\bm S}_\vi \cdot {\bm S}_\vj \right\rangle \sim Q_{f,\vi \vj} = Q_{f,\vj\vi} = - i  \left[   \psi_{\vi}^\dagger e_{\vi\vj }^{\vphantom\dagger} U_{\vi \vj}^{\vphantom\dagger} \psi_{\vj }^{\vphantom\dagger} + \vi \leftrightarrow \vj \right]
\,. \label{VBSL}
\end{align}
In the cuprates, modulations of $Q_{f,\vi\vj}$ would show up as modulations in the charge density on the sites (and similarly for modulations in $Q_{b,\vi\vj}$ below).

Turning to the bosonic partons, and following Ref.~\onlinecite{PNAS_pseudo}, we can also write down the most general effective Lagrangian for the $B_\vi$, keeping only terms quadratic and quartic in the $B_\vi$, and with only on-site or nearest-neighbor couplings:
\beq
\mathcal{L}(B)  = \sum_\vi \left| D_\tau B_\vi \right|^2 + r \sum_{\vi} B^\dagger_{\vi} B_{\vi}^{\vphantom\dagger}   - i w_1 \sum_{ \langle \vi\vj \rangle }   \left[B_{\vi}^\dagger e_{\vi\vj }^{\vphantom\dagger} U_{\vi \vj}^{\vphantom\dagger} B_{\vj }^{\vphantom\dagger}  + \vi \leftrightarrow \vj \right]
+\mathcal{V}(B)\,. \label{LB00}
\eeq
A linear time derivative term is allowed only in the absence of particle hole symmetry, and so has been omitted.
The couplings 
$r$, $w_1$ are real Landau parameters, and the quartic terms are in $\mathcal{V}(B)$.
These quartic terms are more conveniently expressed in terms of quadratic gauge invariant observables. By examining the transformations in Table~\ref{PSG1}, we can deduce the following correspondences between bilinears of the $B$ with those of the bilinears of the gauge-neutral electrons:
\begin{align}
&\mbox{site charge density:~}\left\langle c_{\vi \alpha}^\dagger c_{\vi \alpha}^{\vphantom\dagger} \right\rangle \sim \rho_{\vi} \equiv B^\dagger_\vi B_\vi^{\vphantom\dagger} \nonumber \\
&~~~~~~~~~~~~~~~~~~~~~~~\mbox{ (the correspondence between $\rho_\vi$ and site charge density holds} \nonumber \\ 
&~~~~~~~~~~~~~~~~~~~~~~~~~~~~~\mbox{only in the absence of particle-hole symmetry; see Section~\ref{sec:spin-gap}),} \nonumber\\
&\mbox{bond density:~} \left\langle c_{\vi \alpha}^\dagger c_{\vj \alpha}^{\vphantom\dagger} + c_{\vj \alpha}^\dagger c_{\vi \alpha}^{\vphantom\dagger} \right\rangle 
\sim Q_{b,\vi \vj} = Q_{b,\vj\vi} \equiv \mbox{Im} \left(B^\dagger_\vi e_{\vi \vj}^{\vphantom\dagger} U_{\vi\vj}^{\vphantom\dagger} B_\vj \right), \nonumber \\
&\mbox{bond current:~} i\left\langle c_{\vi \alpha}^\dagger c_{\vj \alpha}^{\vphantom\dagger} - c_{\vj \alpha}^\dagger c_{\vi \alpha}^{\vphantom\dagger} \right\rangle 
\sim J_{\vi \vj} = - J_{\vj\vi} \equiv  \mbox{Re} \left( B^\dagger_\vi e_{\vi \vj}^{\vphantom\dagger} U_{\vi \vj}^{\vphantom\dagger} B_\vj^{\vphantom\dagger} \right), \nonumber \\
&\mbox{pairing:~} \left\langle \varepsilon_{\alpha\beta} c_{i \alpha} c_{j \beta} \right\rangle \sim \Delta_{\vi \vj} = \Delta_{\vj \vi} \equiv \varepsilon_{ab} B_{a\vi} e_{\vi \vj} U_{\vi \vj} B_{b\vj}\,. \label{sitebond}
\end{align}
Note that the bond density observable $Q_{b,\vi \vj}$ of bosons above has the same symmetry signature as the bond energy $Q_{f,\vi \vj}$ of fermions in (\ref{VBSL}), and both are identical to the hopping terms in $\mathcal{L}(B)$ and $\mathcal{L}(\psi)$ respectively. 
Now we can write an expression for $\mathcal{V}(B)$ by keeping all quartic terms which involve nearest-neighbor sites:
\bea 
\mathcal{V}(B) &=& \frac{u}{2} \sum_{\vi} \rho_{\vi}^2 + V_1 \sum_{\vi} \rho_{\vi} \left( \rho_{\vi + \hat{\bm x}} + \rho_{\vi + \hat{\bm y}} \right) +
g \sum_{\langle \vi \vj \rangle} \left| \Delta_{\vi\vj} \right|^2  \nonumber \\
&~&~~~
 + J_1 \sum_{\langle \vi \vj \rangle}  Q_{b,\vi\vj}^2 + K_1 \sum_{\langle \vi \vj \rangle}  J_{\vi\vj}^2.
\label{fb1}
\eea

We also have the usual flux energy term of lattice gauge theory for the gauge field $U_{\vi \vj}$
\beq
\mathcal{L}(U) = -\frac{1}{g} \sum_{\vi, \vj, \vk, \vl \in \square} \mbox{Tr} \left[ U_{\vi \vj} U_{\vj \vk} U_{\vk \vl} U_{\vl\vi} \right] + \mbox{c.c.}\,,
\eeq
along with a gauge field kinetic energy \cite{Kogut_QCD_RMP}.

Finally, we can consider quartic terms which couple the spinons and chargons directly. From the composite operators defined above we can write down the following terms involving only nearest-neighbor sites
\beq
\mathcal{L}(B\psi) = \sum_{\langle \vi \vj \rangle} \left[ \lambda_1 \, c_{\vi \alpha}^\dagger c_{\vj \alpha}^\pdagger + \lambda_1 \, c_{\vj \alpha}^\dagger c_{\vi \alpha}^\pdagger + \lambda_2 \, Q_{b,\vi \vj}\, Q_{f,\vi \vj} \right]\,.
\label{eq:quarticCouplings}
\eeq

Our aim is to determine the phase diagram of the above square lattice gauge theory as a function of the boson `mass' tuning parameter $r$, and the various quartic boson couplings in (\ref{fb1}). The general physics is that of a transition between Higgs and confining phases of the SU(2) gauge theory, with deconfined conformal gauge theories describing continuous transitions between the phases. When $r$ is large and positive, $B$ excitations are gapped, and we can work with the fermion-only theory in (\ref{Hs})---this theory is expected to confine into an insulator with either N\'eel or VBS order \cite{DQCP3,Sandvik20,Zhou:2023qfi}. On the other hand, when $r$ is negative, $B$ condenses in Higgs phases, and fully quenches the SU(2) gauge field. The Higgs phases break one or more of the global symmetries, based upon the correspondence in (\ref{sitebond}).

\section{Quantum field theory and order parameters}
\label{sec:introqft}

Now we take the continuum limit of the square lattice gauge theory action in Section~\ref{sec:lgt}, and obtain the quantum field theory studied in the present paper. We will take the simplest case in which the boson hopping terms are only nearest-neighbor, as in (\ref{LB00}), so there are only two valleys in the boson dispersion. This will lead to a SU(2) gauge theory with $N_f = 2$ flavors of SU(2) fundamental Dirac fermions $\psi$, and $N_b=2$ flavors of SU(2) fundamental bosons $B$. As for the lattice gauge theory in Section~\ref{sec:lgt}, almost everything follows from the symmetry transformations of the fields: the continuum limits of the transformations in Table~\ref{PSG1} are presented in Table~\ref{tab:continuumSymmetries}.

For the continuum limit action of the fermionic spinons, we follow the notation of Ref.~\cite{Shackleton:2022zzm}, which follows that of earlier related works \cite{DQCP3,Thomson:2017ros,Shackleton:2021fdh}, in obtaining from (\ref{Hs}) the fermionic Lagrangian 
\begin{align}
\mathcal{L}_\psi = i \bar{\psi} \gamma^\mu \left( \partial_\mu - i A^\alpha_{\mu}\sigma^\alpha \right) \psi, \label{LF}
\end{align}
where $\sigma^\alpha$ are the Pauli matrices, $\alpha=x,y,z$, $\gamma^\mu$ are $2\times 2$ Dirac matrices which act on the sublattice space, $A_\mu^\alpha$ is the SU(2) gauge field, and the $\psi$ have an additional $N_f = 2$ valley (`flavor') index which is not shown. From the $\psi$ bilinears, we can make a gauge-invariant 5-component real vector, which represents the $3+2$ components of the N\'eel and VBS order parameters \cite{YingRanThesis,RanWen06,DQCP3}; the N\'eel order is a staggered modulation of the spin in (\ref{SL}), while the VBS order is a modulation of the bond energy in (\ref{VBSL}).
The properties of $\mathcal{L}_\psi$ are invariant under global SO(5)$_f$ rotations of this vector, and all our analysis below will preserve this SO(5)$_f$ symmetry (the $f$ subscript merely denotes that the symmetry acts on the fermions). 

It is a simple matter to generalize (\ref{LF}) to arbitrary integer $N_f$: we allow the valley index to run over $1 \ldots N_f$. After transforming to Majorana fermions, the free fermion Lagrangian has a SO$(4N_f)$ symmetry, and modding out the gauge symmetry as in Ref.~\onlinecite{DQCP3}, we conclude that the Lagrangian $\mathcal{L}_\psi$ has a USp$(2N_f)/\mathbb{Z}_2$ global symmetry.

In the bosonic matter sector, we express the lattice $B_{\vi}$ bosons in terms of complex bosons $B_{as}$, with $a=1,2$ the SU(2) gauge index, and $s=1 \ldots N_b=2$ the valley (`flavor') index \cite{PNAS_pseudo}:
\begin{equation}
B_a(\bm{r})= \left\{ 
\begin{array}{c}
- B_{a1}e^{i\pi (x+y)/2} +B_{a2} (\sqrt{2} + 1) e^{i\pi (x-y)/2}, \\
  \quad \mbox{$x$ even} \\
B_{a1} (\sqrt{2} + 1) e^{i\pi (x+y)/2}  -B_{a2} e^{i\pi (x-y)/2}, \\
 \quad \mbox{$x$ odd} 
\end{array} \right.
\label{eq:eigenmode_NN}
\end{equation}
Under particle-hole symmetry $\mathcal{C}$, the transformations in Table~\ref{PSG1} now imply that $B_{as} \rightarrow B_{as}^\ast$.
Then (\ref{sitebond}) leads to the following gauge-invariant order parameters in the continuum limit \cite{PNAS_pseudo}
\begin{align}
\mbox{$d$-wave superconductor} &:~~ \varepsilon_{ab} B_{a 1} B_{b 2} \nonumber \\
\mbox{$x$-CDW} &:~~  B_{a1}^\ast B_{a1}^\pdagger-  B_{a2}^\ast B_{a2}^\pdagger \equiv B^\dagger \mu^z B^\pdagger \nonumber  \\
\mbox{$y$-CDW}&:~~ B_{a1}^\ast B_{a2}^\pdagger +  B_{a2}^\ast B_{a1}^\pdagger \equiv B^\dagger \mu^x B^\pdagger  \nonumber \\
\mbox{$d$-density wave}&:~~ i \left( B_{a1}^\ast B_{a2}^\pdagger -  B_{a2}^\ast B_{a1}^\pdagger\right) \equiv -B^\dagger \mu^y B^\pdagger  
\label{allorders}
\end{align}
where $\mu$ acts on valley indices. In terms of the lattice order parameters in (\ref{sitebond}), the $d$-wave superconductor has $\Delta_{\vi, \vi+ \hat{\bm x}} = -\Delta_{\vi, \vi+ \hat{\bm y}}$, but is independent of $\vi$.
The charge density waves (CDWs) have period 2 modulations of $Q_{b,\vi\vj}$ and $\rho_\vi$ (the modulations of $\rho_\vi$ are absent when there is particle-hole symmetry, see Section~\ref{sec:spin-gap}), and are site-centered unlike the bond-centered modulations of $Q_{f,\vi\vj}$ in the VBS state. The $d$-density wave order is odd under time-reversal, and has a staggered pattern of electrical currents $J_{\vi \vj}$. 
Note that the CDW and $d$-density wave orders can be written as a SO(3) vector 
$B^\dagger \mu^i B^\pdagger$, $i=x, y, z$.
In combination with the complex superconducting order, the order parameters in (\ref{allorders}) form a SO(5)$_b$ vector, for reasons very similar to the fermions (again the $b$ subscript denotes that this SO(5) acts on the bosons). Computing the magnitude of this SO(5)$_b$ vector, we obtain an important identity which is easily verified by explicit evaluation
\begin{align}
    (B^\dagger B)^2 = \left( B^\dagger \mu^i B \right)^2 + 4 \left| \varepsilon_{ab} B_{a 1} B_{b 2} \right|^2\,. \label{identity}
\end{align}

The continuum limit of the Lagrangian (\ref{LB00}) for the bosonic sector is
\begin{align}
    \mathcal{L}_B & = \left| \left( \partial_\mu - i A_\mu^\alpha \sigma^\alpha \right) B \right|^2 + r |B|^2 + \bar{u} |B|^4  \nonumber \\
    & + v_1 \left( B^\dagger \mu^z B \right)^2 + v_1 \left(   B^\dagger \mu^x B \right)^2 \nonumber \\& 
    + v_2 \left(  B^\dagger \mu^y B \right)^2 + v_3 \left|  \varepsilon_{ab} B_{a 1} B_{b 2} \right|^2 \,. \label{LB}
\end{align}
The first three terms in $\mathcal{L}_B$ have the SO(5)$_b$ global symmetry, for reasons essentially identical to those for $\mathcal{L}_f$. All the order parameters in (\ref{allorders}) are degenerate in this limit. This degeneracy and the SO(5)$_b$ symmetry are broken by the $v_{1,2,3}$ terms in (\ref{LB}), which are simply squares of the order parameters in (\ref{allorders}). The identity in (\ref{allorders}) was overlooked in Ref.~\cite{PNAS_pseudo}, and has the consequence that the 5 quartic terms in (\ref{LB}) are not all independent---this has no material consequence to the mean-field results of Ref.~\cite{PNAS_pseudo}, apart from a redundant labeling of couplings.
In the Higgs phase where $B$ is condensed, one of the order parameters in (\ref{allorders}) must be non-zero, and, in mean-field theory, the choice is determined by the relative values of $v_{1,2,3}$ \cite{PNAS_pseudo}.

The generalization of the first three terms in (\ref{LB}) to arbitrary integer $N_b \geq 2$, $N_b$ even 
is straightforward, but the $v_{1,2,3}$ terms in require further consideration. 
We limit ourselves to the case $v_1=v_2$, so that the CDW orders and the $d$-density wave orders become degenerate. Then we can write (\ref{LB}) as
\begin{align}
    \mathcal{L}_B & = \left| \left( \partial_\mu - i A_\mu^\alpha \sigma^\alpha \right) B \right|^2 + r |B|^2 + \bar{u} |B|^4  \nonumber \\
    & + v_1 \left(B^\dagger \mu^i B \right)^2 + v_3 \left|  \varepsilon_{ab} B_{a 1} B_{b 2} \right|^2 \,. \label{LB1}
\end{align}
Next, we use the redundancy implied by (\ref{identity}) to set $v_1=0$ in (\ref{LB1}). Then one extension of (\ref{LB1}) to general $N_b$ for the bosonic flavor indices is obtained by replacing $\varepsilon_{st}$ in the $v_3$ term by $\mathcal{J}_{st}$ the USp($N_b$) invariant tensor, consisting of $N_b/2$ copies of $\varepsilon_{st}$ along the diagonal. (An alternative large $N_b$ extension in which $v_3$ is set to zero is discussed in Appendix~\ref{app:alternative}.)
In this manner we obtain a Lagrangian valid for any $N_b$ (following conventions in Ref.~\cite{Podolsky:2012pv})
\begin{align}
\mathcal{L}_B =  \left| \left( \partial_\mu - i A_\mu^\alpha \sigma^\alpha \right) B \right|^2  +   \frac{u}{2 N_b} \, (|B_{as}|^2 - N_b/g)^2  -
\frac{v}{N_b} \, \left| B^T \mathcal{J} \varepsilon  B \right|^2 \,. \label{e2}
\end{align}
Recall that the indices $a,b$ act on the SU(2) gauge indices, and not the flavor indices, and so do not need a large $N_b$ generalization. 
For $N_b=2$, the correspondence to the couplings in (\ref{LB}) is $u = 2 N_b \bar{u}$, $g = -u/r$, $v=-N_b v_3/4$. 
For general $N_b$, the order parameters in (\ref{allorders}) are replaced by the SU(2) gauge-invariant operators
\begin{align}
\mbox{$d$-wave superconductor} &:~~ \mathcal{J}_{st} \varepsilon_{ab} B_{a s} B_{b t} \nonumber \\
\mbox{charge order} &:~~  B_{as}^\ast T^{i}_{st} B_{at}^\pdagger
\label{allordersN}
\end{align}
where $T^i$ are generators of USp($N_b$)
obeying
\begin{align}
T^{i \dagger} = T^i \quad, \quad T^{iT} \mathcal{J} + \mathcal{J} T^i = 0\,.
\label{eq:constraints}
\end{align}
We refer to the combined and degenerate CDW and $d$-density orders simply as `charge order'.

We can now use standard methods to generate a large $N_b$ expansion of (\ref{e2}) at fixed $u$, $g$, and $v$. The coupling $g$ will be used to tune across the transition, while $v$ will determine the fate of Higgs phase where $B$ is condensed.
The theory in (\ref{e2}) has a global USp($N_b$)$\times$U(1) symmetry, and the Higgs phase with $B$ condensed either breaks the U(1) symmetry leading to $d$-wave superconductivity,
or breaks the USp($N_b$) symmetry leading to degenerate CDW/$d$-density wave orders. 

At $v=0$, the global symmetry of (\ref{e2}) is enhanced to USp$(2N_b)/\mathbb{Z}_2$ (as for the fermionic spinons \cite{DQCP3}), and the superconducting and charge orders all become degenerate. The enhanced symmetry is evident in the matrix form of the bosonic fields in (\ref{Bdef}), which generalizes in the continuum to
\begin{align}
\mathcal{B}_s = \left( \begin{array}{cc} B_{1s} & - B_{2s}^\ast \\ B_{2s} & B_{1s}^\ast \end{array} \right), \label{Bdef2}
\end{align}
obeying the reality condition
\begin{align}
    \mathcal{B}_s = \sigma^y \mathcal{B}_s^\ast \sigma^y\,. \label{reality}
\end{align}
The USp$(2 N_b)$ global symmetry $U_g$ then acts as right multiplication $\mathcal{B} \rightarrow \mathcal{B} U_g$, where $U_g$ is a $2 N_b \times 2 N_b$ matrix acting on both the $s$ flavor index, and the right matrix index of (\ref{Bdef2}). The condition (\ref{reality}) leads to the defining conditions for USp$(2 N_b)$:
\begin{align}
    U_g^\dagger U_g = 1 \quad, \quad U_g^T \sigma^y U_g = \sigma^y\,.
\end{align}
Note, also, that the SU(2) gauge symmetry in (\ref{gaugedef}) acts a left multiplication $\mathcal{B}_s \rightarrow U \mathcal{B}_s$. As in the fermion case, the USp$(2N_b)$ and gauge SU$(2)$ share a common $\mathbb{Z}_2$ center, and hence the global symmetry is USp$(2N_b)/ \mathbb{Z}_2$.

\begin{table}[htp]
    \centering
    \begin{tabular}{c|c|c}
        Symmetry & $B_{a}$ & $X_{ab}$ \\ \hline
         $T_x$ & $-i \mu^x B_a$ & $\mu^x X_{ab}$ \\
         $T_y$ & $-i \mu^z B_a$ & $\mu^z X_{ab}$  \\
         $P_x$ & $B_a$ & $-i \gamma^x \mu^z X_{ab}$  \\
         $R_{\pi/2}$ &$-\frac{\mu^x + \mu^z}{\sqrt{2}}B_a$& $e^{i\pi \gamma^0 / 4} e^{-i\pi \mu^y / 4} X_{ab}$ 
         \\
         $\mathcal{T}$ & $B$ & $\gamma^0 \mu^y X^*$ \\
         $\mathcal{C}$ & $B^*$ & $X \sigma^y$ \\
         $\text{U}(1)_c$ & $e^{i\theta} B_a$ & $X_{ab}$  \\
         $\text{SU}(2)_g$ & $U_g B$ & $X U_g^\dagger$  \\
         $\text{SU}(2)_s$ & $B$ & $U_s X$  \\
    \end{tabular}
    \caption{We tabulate the action of the microscopic symmetries, along with the $\text{SU}(2)$ gauge transformations, on the continuum fields. To concisely express the action of $\text{SU}(2)$ spin rotation symmetry, we represent the spinon degrees of freedom in terms of a matrix of Majorana fermions $X$. The $\gamma$ matrix $\gamma^0$ is the labels the temporal component.}
    \label{tab:continuumSymmetries}
\end{table}

The full action of the microscopic symmetries on the continuum fields is listed in Table~\ref{tab:continuumSymmetries}. To retain a concise representation of the $\text{SU}(2)$ spin rotation symmetry, we re-express our spinon degrees of freedom in terms of Majorana fermions. Following Ref~\onlinecite{DQCP3}, we introduce the $4 \times 2$ matrix of Majorana fermions $X_{a, s; b}$. Here $a, s, b$ are the spin, valley and gauge indices, respectively. The relation between $X$ and the Dirac fermions is given by
$\psi_{a, s} = i \sigma^y_{a, b} X_{1, s, b}$.  The $\text{SU}(2)$ gauge symmetry acts as $X_{a, s; b}  \rightarrow X_{a, s; c} U^\dagger_{cb}$ and $\text{SU}(2)$ spin rotation symmetry acts as $X_{a, s; b}  \rightarrow U_{ac} X_{c, s; b}$. The action of all the symmetries apart from spin rotation symmetry lifts directly to the complex fermions, although a $\text{U}(1)$ subgroup corresponds to a uniform phase rotation $\psi \rightarrow e^{i \theta} \psi$. Both representations will be utilized here - the Majorana representation for when a complete symmetry analysis is required, and the Dirac representation for perturbative computations.

Along with the gauge-invariant fermion and boson bilinears noted above, we will also consider mixed gauge-invariant bilinears which lead to the electron operator measured in photoemission experiments. The quantum field theory yields the electron operator near the 4 nodal points ${\bm k} = (\pm \pi/2, \pi/2)$. The particular combination of low-energy spinons and chargons that correspond to these nodal excitations is rather complicated, as the spinor structure of the Dirac spinons must be unpacked, i.e we consider the fields $\psi_{as\alpha}$ with gauge index $a$, valley index $s$, and spinor index $\alpha$ (which microscopically corresponds to a sublattice index). Suppressing the valley index and taking the Pauli matrices $\mu^i$ to act on both chargon and spinon valley indices, $B_a^* \mu^i \psi_{a \alpha} \equiv B_{as}^* \mu^i_{s t} \psi_{at \alpha}$, we have
\begin{equation}
\begin{aligned}
  C_{{\bm k} = ( \pi/2, \pi/2)} &\propto \begin{pmatrix}  B_{a}^* i\mu^y \left( \psi_{a 1} + \left( \sqrt{2} + 1 \right) \psi_{a2} \right), \\ 
  \epsilon_{ab} B_a  \left( \left( \sqrt{2} + 1 \right) \psi_{a 1} - \psi_{a 2} \right) 
  \end{pmatrix}  \\
C_{{\bm k} = ( -\pi/2, \pi/2)}  &\propto \begin{pmatrix} - B_{a}^* i\mu^z \left( \left( \sqrt{2} + 1 \right) \psi_{a 1} + \psi_{a2} \right) \\ 
  \epsilon_{ab} B_a \mu^x \left(-\psi_{a 1} + \left( \sqrt{2} + 1 \right) \psi_{a 2} \right). 
  \end{pmatrix}  \\
\label{cBpsi}
\end{aligned}
\end{equation}

As we will show, generic operators of the form $B^*_{as} \psi_{as'\alpha}$ and $\epsilon_{ab} B_{as} \psi_{b s' \alpha}$ are all renormalized in the same way at criticality, so the details of Eq.~\ref{cBpsi} will not be relevant for computing the scaling dimension of the electron operator.

We will analyze the theory $\mathcal{L}_\psi + \mathcal{L}_B$ in (\ref{LF}) and (\ref{e2}) in the limit of large $N_f$ and $N_b$, with a fixed ratio $N_f/N_b$. 
We obtain the leading $1/N_{f,b}$ corrections to the scaling dimensions of the gauge-invariant fermion and boson bilinear order parameters, and also the electron operators in (\ref{cBpsi}). We will also obtain the corresponding properties in an alternative large $N_b$ limit in Appendix~\ref{app:alternative}.

\section{Fermion-boson interactions and spin gaps}
\label{sec:spin-gap}

In Section~\ref{sec:introqft}, we constructed a Lagrangian describing spinon and chargon fluctuations and their coupling to a shared $\text{SU}(2)$ gauge field. Importantly, there exist three independent quartic chargon interactions which are relevant at tree-level and must be tuned in order to reach a continuous transition. In this section, we consider symmetry-allowed interactions between the spinons and chargons. The reason for this is twofold. First, quartic interactions involving two spinons and two chargons are marginal at tree-level, and corrections to their scaling dimension are important for the behavior of the critical theory. 
Second, condensation of the chargons can qualitatively modify the dispersion of the spinons in the charge-ordered phase, either by producing a gap or generating a Fermi surface. Note that upon condensation of the chargons, the spinon becomes associated with the electron, and these dispersion modifications are reflected in the electronic spectral function. We show that in fact \textit{no} quartic chargon-spinon interactions are allowed by the microscopic symmetries in the critical theory, provided we enforce particle-hole symmetry. Relaxing particle-hole symmetry admits two quartic interactions. In the charge ordered phase, these terms shift the Fermi energy of the Dirac spinons, thereby inducing a spinon Fermi surface.

In this section, we will use the Majorana representation of the fermionic spinons; the explicit action of spin rotation symmetry is essential in our symmetry analysis. In this language, a generic quartic interaction that respects both charge conservation and spin rotation invariance can be expressed in the form
\begin{equation}
    \begin{aligned}
        \sum_{\alpha, \beta, j} A_{\alpha, \beta, j} \tr \left[ B \mu^\alpha B^\dagger \overline{X} \gamma^j\mu^\beta X \right],
    \end{aligned}
\end{equation}
where $\bar{X} \equiv X^\dagger \gamma^0$ and $A$ is a coefficient tensor, not to be confused with the gauge field. The indices $\alpha\,, \beta\,, j$ run over four variables, the three Pauli and $\gamma$ matrices as well as an additional identity element. We perform a systematic search for symmetry-allowed quartic couplings by deducing the action of the microscopic symmetries on $A_{\alpha, \beta, j}$, which we regard as a $4^3 = 64$-dimensional vector. Symmetry-allowed quartic terms are given by choices of $A$ which have eigenvalue $1$ under all the symmetries, the existence of which can be checked numerically.

With this approach, we deduce two terms that are allowed by all the microscopic symmetries, but are odd under particle-hole symmetry which we assume to be emergent in the critical theory:
\begin{equation}
    \begin{aligned}
        &\tr \left[ B B^\dagger \overline{X} \gamma^0 X\right]\,,
        \\
        &\tr \left[ B \mu^z B^\dagger \overline{X} \mu^z \gamma^x X\right]+\tr \left[ B \mu^x B^\dagger \overline{X} \mu^x \gamma^y X\right]\,.
        \label{eq:continuumQuartic}
    \end{aligned}
\end{equation}
 One can also consider analogous quartic couplings of the form $\sum_{\alpha, \beta, j} C_{\alpha, \beta, j} \tr \left[ B \mu^\alpha B^\dagger \right] \tr \left[ \overline{X} \gamma^j\mu^\beta X \right]$. The tensor $C$ transforms identically to $A$; however, the two quartic couplings in this case vanish identically due to the anticommutation relations of the Majorana fermions. These results are consistent with taking the continuum limit of the quartic spinon-chargon interactions on the lattice given by (\ref{eq:quarticCouplings}), where we find that the leading order terms with no derivatives vanish. Allowing for quartic interactions that break particle-hole symmetry, such as an on-site chemical potential or a second-neighbor electron hopping, generate the continuum interactions in (\ref{eq:continuumQuartic}). The first term acts as a chemical potential and, at each of the two gapless points in momentum space, induces an equal and opposite shift in the Fermi energy on the two species of spinons.

 Quartic interactions do not generate a spin gap in the ordered phases. To find six-term interactions that can open up a spin gap in the CDW phase, we take the approach of considering the CDW order parameter, $B^\dagger \mu^z B$ and $B^\dagger \mu^x B$ for $x$-CDW and $y$-CDW respectively, and coupling them to a quartic chargon-spinon interaction that has the same symmetry transformations. Multiple six-term interactions can be obtained in this manner; however, only two are capable of producing a spin gap, which are
\begin{equation}
   \begin{aligned}
        &B^\dagger \mu^z B \tr \left[ B \mu^x B^\dagger \overline{X} \mu^y X\right]\,,
        \\
        &B^\dagger \mu^x B \tr \left[ B \mu^z B^\dagger \overline{X} \mu^y X\right]\,.
    \end{aligned}
\end{equation}
Note that these terms vanish unless \textit{both} the $x$-CDW and $y$-CDW terms are non-zero. This is consistent with the fact that, once we are in the CDW phase, one is allowed to add non-gauge-invariant terms to the spinon dispersion which break translational symmetry. The symmetry transformations of gauge singlet and triplet spinon bilinears were tabulated in Ref.~\onlinecite{Thomson:2017ros}; from this analysis, one can conclude that the only possible mass term in the CDW phase, $\tr \left[ \sigma^a \overline{X} \mu^y X\right]$, must be odd under translations in both the $x$ and $y$ directions. This term also breaks particle-hole symmetry; however, as it is proportional to four powers of the chargon condensate, it will generically be smaller than the previously-discussed perturbations which generate a spinon Fermi surface.

\section{Large $N_b$ saddle point}
\label{sec:saddlepoint}

This section examines the bosonic theory $\mathcal{L}_B$ in (\ref{e2}), and determines its phase diagram at $N_b = \infty$. We introduce decouplings fields $\lambda$ and $\Delta$ to obtain from (\ref{e2})
\begin{align}
\mathcal{L}_B = |D_\mu B_{as}|^2 + \frac{N_b \lambda^2}{2u}  + \frac{N_b |\Delta|^2}{v} + \mathrm{i} \lambda ( |B_{as}|^2 - N_b/g ) -  \Delta \, \mathcal{J}_{st} \varepsilon_{ab}  B^\ast_{as} B^\ast_{bt}  -  \Delta^\ast \, \mathcal{J}_{st} \varepsilon_{ab} B_{bt} B_{as}. \label{LBa}
\end{align}
The saddle point value of $\i \lambda$ will determine the mass of the $B$ bosons, while $(N_b/v) \Delta$ is the superconducting order parameter in (\ref{allordersN}).
In order to carry out the Gaussian integral over the $B$ bosons, it is convenient to define a Nambu basis for $B$. We would like the quadratic terms in $B$ which are associated with pairing to be completely off diagonal in our choice of basis and for the rest of the terms to be diagonal. We therefore use the fact that $\mathcal{J}_{st}$ is anti-symmetric to construct the Nambu basis:
\begin{equation}
\mathcal{B}_{m}=
    \begin{pmatrix}
        B_{1 ,2m-1}\\B_{2,2m-1}\\B^*_{2,2m}\\-B^*_{1,2m}
    \end{pmatrix}
    \label{bvec1}
\end{equation}
Here we have used $1,2$ to label the indices corresponding to the SU(2) gauge symmetry and $m=1,...,{N_b}/{2}$.

After integrating out the bosons, the effective action is:
\begin{equation}\label{Integratedout}
    S_{\text{eff.}}= \frac{N_b}{2} \text{Tr}\left[\text{ln}(\mathcal{G}^{-1})\right]+\frac{N_b\lambda^2}{2u}+\frac{N_b |\Delta|^2}{v}-\mathrm{i}\lambda\frac{N_b}{g},
\end{equation}
where
\begin{equation}
    \mathcal{G}^{-1}=\begin{pmatrix}
        \mathrm{i}\lambda-(\partial_{\mu}+\mathrm{i}A_{\mu}^j\sigma_j)^2 & -2\Delta\\ -2\Delta^* & \mathrm{i}\lambda-(\partial_{\mu}+\mathrm{i}A_{\mu}^j\sigma_j)^2
    \end{pmatrix}
\end{equation}
is a $4 \times 4$ matrix. We assume $A_\mu^j = 0$ at the saddle point (preserving gauge and Lorentz symmetry). 
The saddle point equation for $\lambda$ is
\begin{equation}\label{LambdaSP}
    \frac{\mathrm{i}\lambda N_b}{u}+\frac{N_b}{g}=\int\frac{d^3p}{(2\pi)^3}\left(\mathrm{i}\lambda+p^2\right)\frac{2N_b}{(\mathrm{i}\lambda+p^2)^2-4|\Delta|^2}\,,
\end{equation}
and that for $\Delta$ is
\begin{equation}\label{DeltaSP}
    \frac{N_b}{v}=\int\frac{d^3p}{(2\pi)^3}\frac{4N_b}{(\mathrm{i}\lambda+p^2)^2-4|\Delta|^2} \,.
\end{equation}
At the saddle point where $\Delta=0$, setting $\bar{\lambda}\equiv\mathrm{i}\lambda$ we recover the result of Ref.~\onlinecite{KaulSS08}.
 \begin{equation}
    \int\frac{d^3p}{(2\pi)^3}\frac{1}{(\bar{\lambda}+p^2)}=\frac{\bar{\lambda} }{2u}+\frac{1}{2g}\,.
\end{equation}
In what follows, we will always assume $g>0$.

\subsection{Solving saddle point equations}

\begin{figure}
    \centering
    \includegraphics[scale=.5]{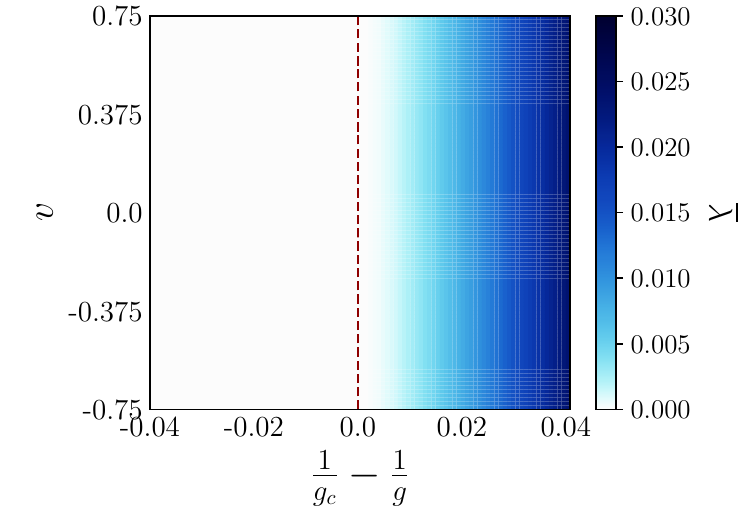}
    \caption{We show the saddle point solutions for $\overline{\lambda}$ as a function of $\frac{1}{g_c}-\frac{1}{g}$ and $v$ for $u=1.5$ for the solution where only $\overline{\lambda}$ is condensed and $B$ and $\Delta$ are both zero. Such a solution only exists when $g>g_c$ and we note the value of $\overline{\lambda}$ has no dependence on $v$ when $\langle\Delta\rangle=0$. The boundary after which $\overline{\lambda}$ is nonzero is denoted with a dotted red line.   }
    \label{fig:onlyLambda}
\end{figure}

In integrating out the $B$ bosons, we have assumed there is no condensate in $B$. We will first solve the saddle point equations under the assumption that $\langle B\rangle=0$ and then consider alternate solutions where $B$ condenses. Under this assumption, the saddle point equations for $\Delta$ and $\overline{\lambda} \equiv \mathrm{i} \lambda$  obtained from integrating (\ref{DeltaSP}) and (\ref{LambdaSP}) are:
\begin{equation}\label{HeisSP}
    \frac{1}{v}=\frac{1}{4\pi}\frac{1}{|\Delta|}\left[\sqrt{\overline{\lambda}+2|\Delta|}-\sqrt{\overline{\lambda}-2|\Delta|}\right]\,,
\end{equation}
\begin{equation}\label{LambdaInt}
    \frac{\overline{\lambda}}{u}+\frac{1}{g}=-\frac{1}{4\pi}\left[\sqrt{\overline{\lambda}+2|\Delta|}+\sqrt{\overline{\lambda}-2|\Delta|}-\frac{4\pi}{g_c}\right]\,,
\end{equation}
where $1/g_c=\Lambda/\pi^2$, with $\Lambda$ the momentum space cutoff. We first note the existence of a solution where $\Delta=0$ and $\overline{\lambda}$ is condensed obtained by neglecting (\ref{DeltaSP}), setting $\Delta=0$ in (\ref{LambdaInt}), and solving (\ref{LambdaInt}) for $\overline{\lambda}$. Such a solution is shown in Fig.~\ref{fig:onlyLambda}.

\begin{figure}
    \centering
    \includegraphics[scale=.5]{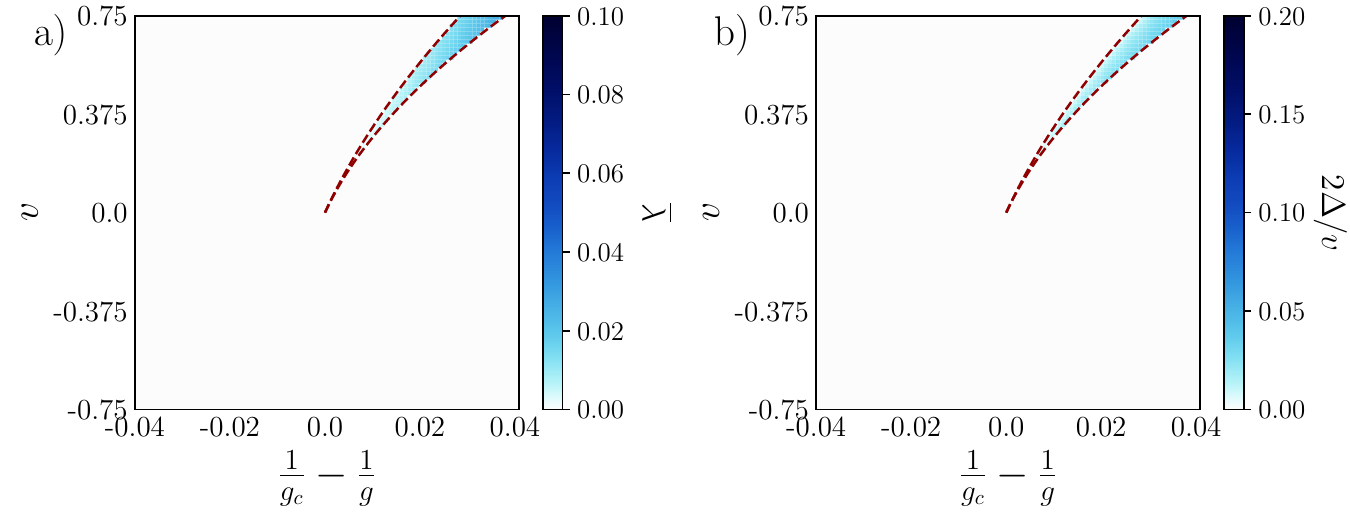}
    \caption{We show the saddle point solutions for $\overline{\lambda}$ (a) and $\Delta$ (b) for the class of solution where $\Delta$ and $\overline{\lambda}$ are both nonzero but $B$ is assumed to not be condensed as a function of $v$ and $\frac{1}{g_c}-\frac{1}{g}$. A real, positive solution for $\overline{\lambda}$ and $|\Delta|$ only exists for the narrow strip shown in the region where $g>g_c$ and $v>0$. On the lower curve of the region of existence of this solution we have $\overline{\lambda}=2|\Delta|$. The boundary enclosing the region where each quantity becomes nonzero is denoted with a dotted red line.  }
    \label{fig:LambdaDeltaCond}
\end{figure}

We can also find solutions of Eq.~\ref{HeisSP} and Eq.~\ref{LambdaInt}  where $\lambda$ and $\Delta$ are both condensed. Multiplying the saddle point equation for $\overline{\lambda}$ and $\Delta$ together yields a constraint on $\overline{\lambda}$ which is independent of $\Delta$:
\begin{equation}
    \overline{\lambda}=u\left(\frac{1}{g_c}-\frac{1}{g}-\frac{v}{4\pi^2}\right)
\end{equation}
We then only need to assume the above relation for $\overline{\lambda}$, substitute this expression into Eq.~\ref{HeisSP} or Eq.~\ref{LambdaInt}, and solve for $\Delta$. The resulting solution is shown in Fig.~\ref{fig:LambdaDeltaCond}, and exists only on a narrow strip for positive $v$ and $\frac{1}{g_c}-\frac{1}{g}$. On the lower boundary of this strip of solution, we have $\overline{\lambda}\rightarrow 2|\Delta|$.

\begin{figure}
    \centering
    \includegraphics[width=\linewidth]{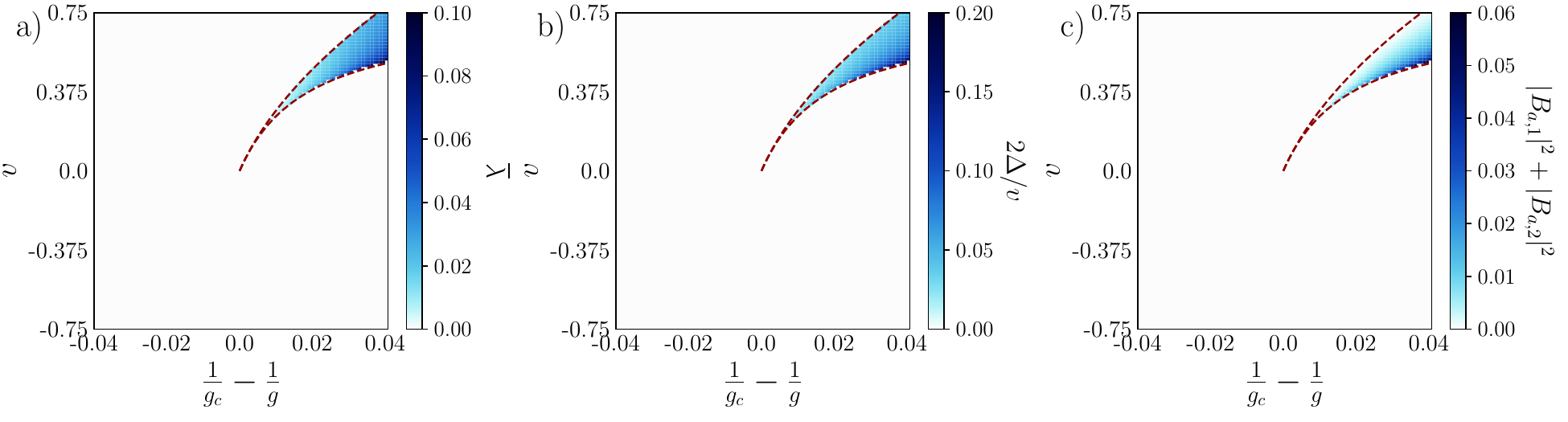}
    \caption{We show the saddle point solutions for $\overline{\lambda}$ (a), $\Delta$ (b), and $B$ (c) for the class of solution where all are allowed condensed for a the first branch of the solution corresponding to Eq.~\ref{SolnB} as a function of $\frac{1}{g_c}-\frac{1}{g}$ and $v$ for $u=1.5$. The boundary enclosing the region where each quantity becomes nonzero is denoted with a dotted red line. We note that a solution with a positive and real $\sqrt{|\Delta|}$ only exists for $v>0$ and $g>g_c$. The upper boundary of this solution aligns with the lower boundary of the solution in Fig.~\ref{fig:LambdaDeltaCond}.}
    \label{fig:LambdaDeltaBbranch1}
\end{figure}

\begin{figure}
    \centering
    \includegraphics[width=\linewidth]{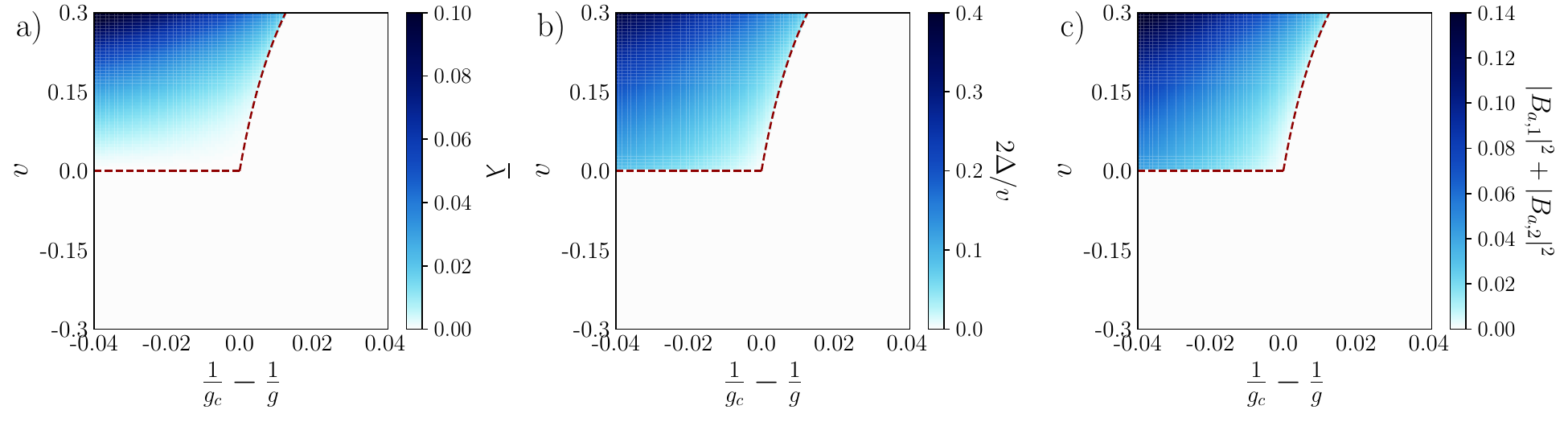}
    \caption{We show the saddle point solutions for $\overline{\lambda}$ (a), $\Delta$ (b), and $B$ (c) for the class of solution where all are allowed condensed for a the second branch of the solution corresponding to Eq.~\ref{SolnB} as a function of $\frac{1}{g_c}-\frac{1}{g}$ and $v$ for $u=1.5$. This class of solution exists only when $v>0$ and unlike the solution corresponding to the first branch of Eq.~\ref{SolnB} plotted in Fig.~\ref{fig:LambdaDeltaBbranch1}, the solution corresponding to to second branch exists when $0<g<g_c$. We note the difference in scale of the magnitude of the plotted quantities as compared to Fig.~\ref{fig:LambdaDeltaBbranch1} for the same range of $v$. The boundary enclosing the region where each quantity becomes nonzero is denoted with a dotted red line.}
    \label{fig:LambdaDeltaBbranch2}
\end{figure}

We now investigate a third class of solution, one where we allow $B$ to condense in addition to $\overline{\lambda}$ and $\Delta$ by allowing for a condensate in the $m=1$ component of (\ref{bvec1})
\begin{equation}
\left\langle \mathcal{B}_{m} \right\rangle =
    \sqrt{N_b} \begin{pmatrix}
        B_{1, 1}\\B_{2, 1}\\B^*_{2, 2}\\-B^*_{1, 2}
    \end{pmatrix} \delta_{m 1}\,.
    \label{bvec2}
\end{equation}
After integrating out the $m>1$ components, we obtain the large $N_b$ effective action generalizing 
(\ref{Integratedout}) 
\begin{align}\label{CondensedSP}
    S_{\text{eff.}}= & \frac{N_b}{2} \text{Tr}\left[\text{ln}(\mathcal{G}^{-1})\right]-\frac{N_b\overline{\lambda}^2}{2u}+\frac{N_b |\Delta|^2}{v}-\overline{\lambda}\frac{N_b}{g} \nonumber \\
    & + N_b \left[ \overline{\lambda} (|B_{a1}|^2+|B_{a2}|^2)-\Delta B^*_{a s}B^*_{b t}\varepsilon_{ab}\varepsilon_{st}-\Delta^* B_{a s}B_{b t}\varepsilon_{ab}\varepsilon_{st} \right]\,.
\end{align}
The saddle point equations for $B$ are:
\begin{equation}
    \overline{\lambda}B^*_{1, 1}-2\Delta^*B_{2, 2}=0 \qquad  \overline{\lambda}B^*_{2, 1}+2\Delta^*B_{1,2}=0 \qquad \overline{\lambda}B^*_{1, 2}+2\Delta^*B_{2, 1}=0 \qquad  \overline{\lambda}B^*_{2, 2}-2\Delta^*B_{1, 1}=0\,.
\end{equation}
We note that combining the above equations produces the constraint:
\begin{equation}
    \overline{\lambda}=2|\Delta|\,.
\end{equation}
Additionally, we note that the saddle point equations for $B$ imply that if $\overline{\lambda}$, $\Delta$, and $B$ all condense, the $d$-wave order parameter in (\ref{allorders}) also must condense. The saddle point equation for $\overline{\lambda}$ becomes:
\begin{equation}
    \frac{2|\Delta|}{u}+\frac{1}{g}-\frac{1}{g_c}-(|B_{a,1}|^2+|B_{a,2}|^2)=-\frac{\sqrt{|\Delta|}}{2\pi}\,,
\end{equation}
while the saddle point equation for $\Delta$ when $B$ is nonzero becomes:
\begin{equation}
    \frac{1}{v}-\frac{1}{|\Delta|}(|B_{a,1}|^2+|B_{a,2}|^2)=\frac{1}{2\pi\sqrt{|\Delta}|}\,.
\end{equation}
Combining the two equations yields:
\begin{equation}
    |B_{a,1}|^2+|B_{a,2}|^2=-\frac{\sqrt{|\Delta|}}{2\pi}+\frac{|\Delta|}{v} \implies |\Delta|\left(\frac{1}{2v}-\frac{1}{u}\right)-\frac{\sqrt{|\Delta|}}{2\pi}+\frac{1}{2}\left(\frac{1}{g_c}-\frac{1}{g}\right)=0\,,
\end{equation}
such that we have solutions corresponding to $\sqrt{|\Delta|}$:
\begin{equation}\label{SolnB}
    \sqrt{|\Delta|}=\frac{\frac{1}{2\pi}\pm\sqrt{\left(\frac{1}{2\pi}\right)^2-2\left(\frac{1}{2v}-\frac{1}{u}\right)\left(\frac{1}{g_c}-\frac{1}{g}\right)}}{2\left(\frac{1}{2v}-\frac{1}{u}\right)}
\end{equation}
The two branches for which $\sqrt{|\Delta|}$ is real and positive are shown in Fig.~\ref{fig:LambdaDeltaBbranch2} and Fig.~\ref{fig:LambdaDeltaBbranch1}.

The first order phase boundary between phases A and B is determined by where the argument of the square root in (\ref{SolnB}) becomes negative and lies along the curve: 
\begin{align}
    v=\left({1}/{g_c}-{1}/{g}\right)\left[(2\pi)^{-2}+({2}/{u})({1}/{g_c}-{1}/{g})\right]^{-1}\,.
\end{align}

Finally, there is a final type of possible solution where only $B$ is condensed, and $\langle\overline{\lambda}\rangle=\langle\Delta\rangle=0$. Such a solution must obey $|B_{1}|^2+|B_{2}|^2=-\frac{1}{g_c}+\frac{1}{g}$, but unlike the solution where $\Delta$ and $\overline{\lambda}$ are also condensed, there is no constraint from the saddle point equations to determine which order parameters in (\ref{allorders}) are nonzero when $\langle B\rangle\neq 0$. We argue that the order parameter which condenses can be determined from the sign of $v$ from the original action in (\ref{e2}) by noting that when $v$ is positive, it is energetically favorable for the superconducting order parameter in the $B$'s to become nonzero while if $v$ is negative, it is favorable for the $d$-density wave or CDW to become nonzero.

\begin{figure}
    \centering
    \includegraphics[width=\linewidth]{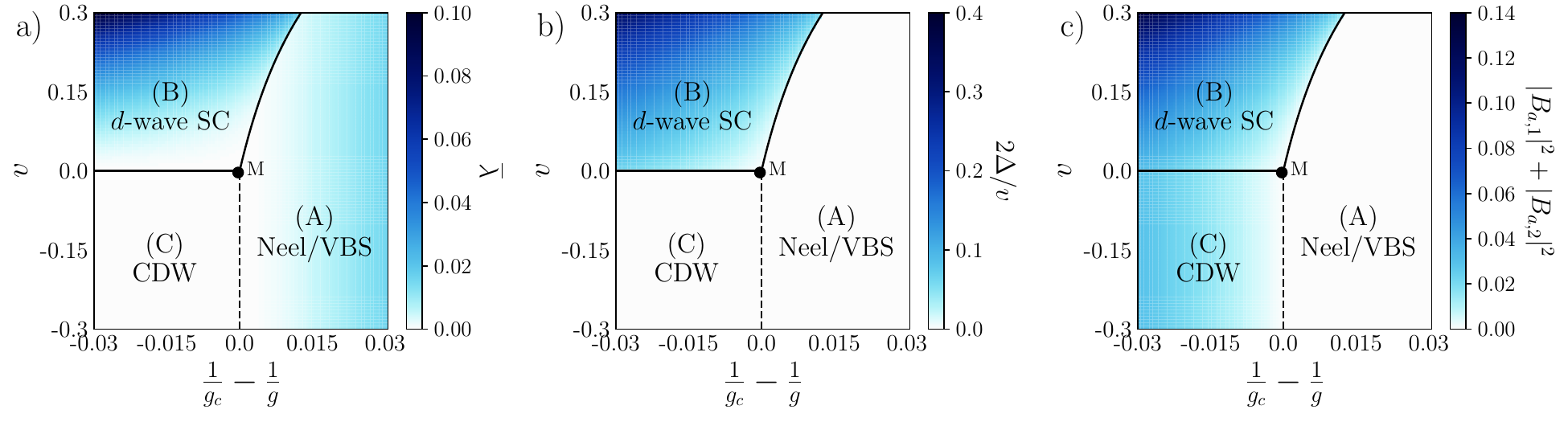}
    \caption{We show the lowest energy saddle point solutions for $\overline{\lambda}$ (a), $|\Delta|$ (b), and $|B_{a,1}|^2+|B_{a,2}|^2$ and (c) as a function of $\frac{1}{g_c}-\frac{1}{g}$ and $v$. We denote the boundaries between each phase with a black solid line if the phase boundary is first order and a black dotted line if the phase boundary is second order. The lowest energy solution for $g>g_c$ is the solution plotted in Fig.~\ref{fig:onlyLambda} with only $\langle\overline{\lambda}\rangle\neq0$ in the region where the solution shown in Fig.~\ref{fig:LambdaDeltaBbranch2} (the one where $\Delta$, $B$, and $\overline{\lambda}$ are all condensed) does not exist; this solution corresponds to either Neel or VBS order since neither $\Delta$ nor $B$ are condensed. In the region where $v>0$ where the solution shown in Fig.~\ref{fig:LambdaDeltaBbranch2} does exist, it is always the lowest energy solution; this solution corresponds to a $d$-wave superconductor. For $0<g<g_c$ and $v<0$, the only possible solution is the one where only $B$ is condensed; based off our arguments in the text, since this solution is the lowest energy only for $v<0$, such a solution corresponds to charge order.}
    \label{fig:PhasesAll}
\end{figure}

We have presented four possible classes of solutions; a solution where only $B$ is condensed, a solution where only $\overline{\lambda}$ is condensed, a solution where $\overline{\lambda}$ and $\Delta$ are condensed but $\langle B\rangle=0$, and a solution where $\overline{\lambda}$, $\Delta$, and $B$ all condense. The phase diagram is then determined by plugging each solution into (\ref{CondensedSP}) and choosing the one with the lowest free energy. After integration, (\ref{CondensedSP}) becomes:
\begin{equation}
\begin{split}
    \frac{S_{\text{eff.}}}{N_b}= & -\frac{1}{6\pi}\left[\left(\overline{\lambda}+2|\Delta|\right)^{3/2}+\left(\overline{\lambda}-2|\Delta|\right)^{3/2}\right]+\frac{2}{3\pi^2}\Lambda\overline{\lambda}-\frac{2}{9\pi^2}\Lambda^3+\frac{1}{6\pi^2}\Lambda^3\text{ln}\left[(\overline{\lambda}+\Lambda^2)^2-4|\Delta|^2\right]-\frac{\overline{\lambda}^2}{2u} \\ &
    +\frac{ |\Delta|^2}{v}-\frac{\overline{\lambda}}{g} 
     +  \overline{\lambda} (|B_{a1}|^2+|B_{a2}|^2)-\Delta B^*_{a s}B^*_{b t}\varepsilon_{ab}\varepsilon_{st}-\Delta^* B_{a s}B_{b t}\varepsilon_{ab}\varepsilon_{st} \,.
\end{split}
\end{equation}
In practice we compute the above with a cutoff $\Lambda=100$, and find the low energy phases shown in Fig.~\ref{fig:PhasesAll}. We note that when $B$ is condensed such that $\overline{\lambda}=2|\Delta|$, there is no direct dependence of the effective action on $B$ in the above.

\section{Computations at order $1/N_{f,b}$}
\label{sec:1N}

For convenience, we present the complete Lagrangian $\mathcal{L} = \mathcal{L}_\psi + \mathcal{L}_B$ in (\ref{LF}) and (\ref{LBa}) for our SU(2) gauge theory. 
\be
\L=i\bar{\psi} \slashed{D}_{\mu}\psi+ |D_{\mu}B_{as}|^2+\frac{N_b \lambda^2}{2u}+\frac{N_b|\Delta|^2}{v}
+\i\lambda(|B_{as}|^2-N_b/g)- \mathcal{J}_{st}\varepsilon_{ab} (\Delta  B_{as}^*B_{bt}^*+\Delta^* B_{bt} B_{as}). \label{Lcomplete}
\ee
The kinetic term for boson should be understood as 
\be
|(D_{\mu}B_{s})_a|^2\equiv (\p_{\mu} B_a^*-\i B_b^* \sigma^j_{ba} A_j)(\p_{\mu} B_a+\i A_j \sigma_{ab}^j B_b).
\ee
We will study (\ref{Lcomplete}) in a large $N_{f,b}$ expansion, with $N_f/N_b$ fixed. This is similar to the method followed in Ref.~\onlinecite{KaulSS08} for a U(1) gauge theory.

\subsection{Multicritical point at $v=0$}
\label{subsec:v=0}
First we consider the multicritical point M in Fig.~\ref{fig:1}, where we can ignore the pairing field $\Delta$ in (\ref{Lcomplete}), and work with a Lagrangian with $\text{USp}(2N_f) \times \text{USp}(2N_b) / \mathbb{Z}_2$ global symmetry:
\be
\L_0=\mathrm{i}\bar{\psi} \slashed{D}_{\mu}\psi+ |D_{\mu}B_{as}|^2+\frac{N_b \lambda^2}{2u}
+\i\lambda(|B_{as}|^2-N_b/g)\,.
\ee
Taking the Fourier transformation and integrating over the bosons and fermions, we write the free energy as
\be
{\mathcal{F}}_0= \tr \ln \mathcal{G}^{-1}_{b} +N_b\left(\frac{\lambda^2}{2u}-\frac{\lambda}{g}\right)+\tr \ln \mathcal{G}^{-1}_f,
\label{eq:F0}
\ee
where $\mathcal{G}^{-1}_b$ is a $2N_b\times 2N_b$ matrix of block-diagonal form $\mathcal{G}^{-1}_b=\left(\begin{matrix}  \mathcal{G}^{-1}_A & 0 \\ 0 & \mathcal{G}^{-1}_D \end{matrix}
\right)$ in the Nambu basis:
\be
\mathcal{G}^{-1}_{A/D}=\mathbbm{1}\left[\delta_{kk'}k^2+\mathrm{i}\lambda (k-k')+\int \frac{d^3q}{(2\pi)^3}A_{\alpha}(q)A^{\alpha}(k-k'-q)\right]
\pm\sigma^{\alpha} \left[(k+k')_{\mu}A_{\alpha}^{\mu}(k-k')\right]. 
\ee
$\mathcal{G}^{-1}_f$ is the corresponding matrix for the fermionic sector
\be
\mathcal{G}^{-1}_f=\gamma^{\mu}[-\delta_{kk'}k_{\mu} \mathbbm{1} +A_{\mu}^{\alpha}(k'-k) \sigma^{\alpha}]. 
\ee
Next we expand near the saddle point by defining the propagator 
\be
G_B(k)=\frac{1}{k^2+\bar{\lambda}},\quad G_{\psi}=\frac{\slashed{k}}{k^2},
\ee
where $\bar{\lambda}=\mathrm{i}\lambda_c$ is real and positive. 
We expand the matrix log to second order, see appendix \ref{app:detail} for details. The leading correction to the free energy can be computed as 
\be
{\mathcal{F}_0^{(1)}}= \frac{1}{2}\int \frac{d^3p}{(2\pi)^3} \Bigg\{
\Pi_{\lambda} (p) 
 \lambda(p)\lambda(-p) +A_{\alpha}^{\mu}(p)\left(\delta_{\mu\nu}-\frac{p_{\mu}p_{\nu}}{p^2}\right)\Pi_A (p) A^{\alpha}_{\nu}(-p)\Bigg\}+N_b\left(\frac{\lambda^2}{2u}-\frac{\lambda}{g}\right),
\label{eq:eff_F0}
\ee
where the kernels are
\be
\begin{split}
& \Pi_{\lambda}(p)=\frac{2N_b}{4\pi p}\arctan \frac{p}{2\sqrt{\bar{\lambda}}},\\
& \Pi_A (p)= 2N_b\left(
\frac{4\bar{\lambda}+p^2}{8p\pi} \arctan \frac{p}{2\sqrt{\bar{\lambda}}}-\frac{\sqrt{\bar{\lambda}}}{4\pi}\right)+N_f \frac{p}{16}.
\end{split}
\label{eq:kernels}
\ee
The dressed propagators can also be read off,
\be
\begin{split}
D^{ij}_{A,\mu\nu}=\frac{\delta_{ij}}{\Pi_{A}}\left(\delta_{\mu\nu}-\zeta \frac{p_{\mu}p_{\nu}}{p^2}\right),\quad D_{\lambda}=\frac{1}{\Pi_{\lambda}}.
\end{split}
\label{eq:propagators}
\ee 
Here $i,j$ are the gauge indices and $\mu, \nu$ are the spacetime indices. For simplicity we introduce the standard notation
$
(A_{\mu})_{aa'}\equiv \sum_i A_{\mu}^i (\sigma^i)_{aa'}. 
$
The propagators then become
\be
\langle (A_{\mu})_{ab} (A_{\nu})_{a'b'}\rangle (q)=(2\delta_{ab'}\delta_{ba'}-\delta_{aa'}\delta_{bb'}) D_{\mu\nu}(q) =\frac{2\delta_{ab'}\delta_{ba'}-\delta_{aa'}\delta_{bb'}}{\Pi_{A}(q)}\left(\delta_{\mu\nu}-\zeta \frac{q_{\mu}q_{\nu}}{q^2}\right),
\ee
Notice that at the critical point, the kernels reduce to
\be
\Pi_\lambda \rightarrow \frac{2N_b}{8p},\quad \Pi_{A,\mu\nu}\rightarrow  (2N_b+N_f)\frac{ p}{16}. 
\ee

\subsubsection{Dressed boson field}

The anomalous dimension of the $B$ field  is
\be
\text{dim}[B_{as}]=\frac{3-2+\eta_B}{2}=\frac{1}{2}+\frac{\eta_B}{2}, 
\ee
where $a$ is the gauge index, and $s$ is the flavor index as usual. The operator is not gauge-invariant on its own. 
The corrections come from the gauge fields and $\lambda$. We draw the corresponding diagrams in \ref{fig:single_B}. 

\begin{figure}
\begin{tikzpicture}[scale=0.7]
\node at (-5,2) {(a)};
\draw[thick,->-=0.5](-4,0) -- (-2,0);
\draw[thick,->-=0.5](-2,0) -- (2,0);
\draw[thick,->-=0.5](2,0) -- (4,0);
\draw[thick,snake it] (-2,-0.5) .. controls +(-0.4,3) and +(0.4,3) .. (2,-0.5);
\filldraw[fill=white,draw=white] (-2.5,-0.03)--(-1.5,-0.03)--(-1.5,-0.5)--(-2.5,-0.5);
\filldraw[fill=white,draw=white] (2.5,-0.03)--(1.5,-0.03)--(1.5,-0.5)--(2.5,-0.5);
\node at (-3,-0.4) {$k$};
\node at (3,-0.4) {$k$};
\node at (0,-0.4) {$k+p$};
\node at (-3,+0.5) {$\textcolor{blue}{a, s}$};
\node at (3,+0.5) {$\textcolor{blue}{a'', s}$};
\node at (0,+0.5) {$\textcolor{blue}{a', s}$};
\node at (0,2.5) {\textcolor{orange}{$A_{\mu}$}};
\node at (0,1.3) {$p$};
\end{tikzpicture}
\quad\quad
\begin{tikzpicture}[scale=0.7]
\node at (-5,2) {(b)};
\draw[thick,->-=0.5](-4,0) -- (-2,0);
\draw[thick,->-=0.5](-2,0) -- (2,0);
\draw[thick,->-=0.5](2,0) -- (4,0);
\draw[thick,dashed,-<-=0.5] (-2,-0.5) .. controls +(-0.4,3.5) and +(0.4,3.5) .. (2,-0.5);
\filldraw[fill=white,draw=white] (-2.5,-0.03)--(-1.5,-0.03)--(-1.5,-0.5)--(-2.5,-0.5);
\filldraw[fill=white,draw=white] (2.5,-0.03)--(1.5,-0.03)--(1.5,-0.5)--(2.5,-0.5);
\node at (-3,-0.4) {$k$};
\node at (3,-0.4) {$k$};
\node at (0,-0.4) {$k+p$};
\node at (-3,+0.5) {$\textcolor{blue}{a, s}$};
\node at (3,+0.5) {$\textcolor{blue}{a, s}$};
\node at (0,+0.5) {$\textcolor{blue}{a, s}$};
\node at (0,2.6) {\textcolor{orange}{$\lambda$}};
\node at (0,1.65) {$p$};
\end{tikzpicture}
\caption{Correction to boson propagators at $v=0$. }
\label{fig:single_B_0}
\end{figure}
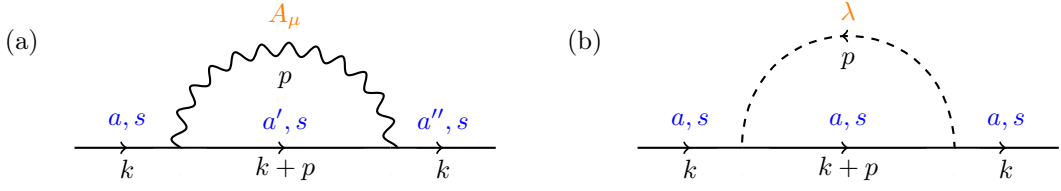

Since the boson propagator is the same for any flavor $(a,s)$, we can first compute the integral and then take care of the indices. The integral corresponding to fig. \ref{fig:single_B_0}(a) is 
\be
\begin{split}
I_{A;1}= & \int \frac{d^3p}{8\pi^3} \left[ G_B(k+p)D_{\mu\nu}(-p)(2k+p)_{\mu}(2k+p)_{\nu}\right] \mid_{\bar{\lambda}=0}\\
\rightarrow \ & -\frac{4}{(2N_b+N_f)\pi^2} \left(\frac{10}{3}+2\zeta \right)k^2\log k, 
\end{split}
\label{eq:I_A_1}
\ee 
where the right arrow means we are extracting the $k^2\log k$ divergence. Computational details for all the integrals can be found in appendix \ref{app:detail}. Taking care of the trace over internal gauge indices, 
\be
\sum_{a',a''} \sigma^j_{aa'}\sigma^j_{a'a''}=\sum_{a',a''} (2\delta_{aa''}-\delta_{a'a}\delta_{a''a})=3.
\ee
The integral for diagram \ref{fig:single_B_0}(b) is
\be
I_{\lambda;1} = \i^2 \int \frac{d^3p}{8\pi^3} \left[G_B(k+p)D_{\lambda}(-p)\right]\big|_{\bar{\lambda}=0}\rightarrow  \frac{2}{3 N_b\pi^2}k^2\log k.
\label{eq:I_lambda_1}
\ee
Summing everything up, we have
\be
\eta_{B}=\frac{2}{3N_b \pi^2}-\frac{12}{(2N_b+N_f)\pi^2} \left(\frac{10}{3}+2\zeta \right).
\label{eq:etaB_0}
\ee

\subsubsection{Charge order parameter}

Next we work out the vertex corrections to get the 
\be
\text{dim}[B^{\dagger}_{as} T^{\alpha}_{st} B_{at}]=2\ \text{dim}[B]+\eta_{\text{vertex}},
\ee
where $T^{\alpha}$ is some generator of the USp$(N_b)$ group that satisfies
\eqref{eq:constraints}. Again we will first do the integrals and then take into account the indices. The relevant diagrams are shown in figure \ref{fig:DW}.

\begin{figure}[htbp]
\centering
\begin{tikzpicture}
\node at (-1,0) {(a)};
\filldraw (0,0) circle (2pt);
\draw[thick,->-=0.5] (0,0)--(1.2,0.75);
\draw[thick,->-=0.5] (1.2,0.75)--(2.4,1.5);
\draw[thick,-<-=0.5] (0,0)--(1.2,-0.75);
\draw[thick,-<-=0.5] (1.2,-0.75)--(2.4,-1.5);
\draw[thick, snake it] (1.2,-0.75)--(1.2,0.75);
\node at (0.35,0.8) {$k_1-p$};
\node at (0.,0.35) {\textcolor{blue}{$a, s$}};
\node at (0.35,-0.8) {$k_2-p$};
\node at (0.,-0.4) {\textcolor{blue}{$a, t$}};
\node at (2,1.7) {$k_1$};
\node at (1.5,1.35) {\textcolor{blue}{$a',s$}};
\node at (2,-1.7) {$k_2$};
\node at (1.5,-1.35) {\textcolor{blue}{$a'',t$}};
\node at (1.6,0.2) {$p$};
\node at (1.6,-0.3) {$\textcolor{orange}{A_{\mu}}$};
\end{tikzpicture}
\quad\quad\quad\quad\quad
\begin{tikzpicture}
\node at (-1,0) {(b)};
\filldraw (0,0) circle (2pt);
\draw[thick,->-=0.5] (0,0)--(1.2,0.75);
\draw[thick,->-=0.5] (1.2,0.75)--(2.4,1.5);
\draw[thick,-<-=0.5] (0,0)--(1.2,-0.75);
\draw[thick,-<-=0.5] (1.2,-0.75)--(2.4,-1.5);
\draw[thick, dashed, ->-=0.5] (1.2,-0.75)--(1.2,0.75);
\node at (0.35,0.8) {$k_1-p$};
\node at (0.,0.35) {\textcolor{blue}{$a, s$}};
\node at (0.35,-0.8) {$k_2-p$};
\node at (0.,-0.4) {\textcolor{blue}{$a, t$}};
\node at (2,1.7) {$k_1$};
\node at (1.5,1.35) {\textcolor{blue}{$a,s$}};
\node at (2,-1.7) {$k_2$};
\node at (1.5,-1.35) {\textcolor{blue}{$a,t$}};
\node at (1.6,0.2) {$p$};
\node at (1.6,-0.3) {$\textcolor{orange}{\lambda}$};
\end{tikzpicture}
\caption{Diagrams contributing to the vertex correction of the density wave scaling dimension at leading order. }
\label{fig:DW_0}
\end{figure}
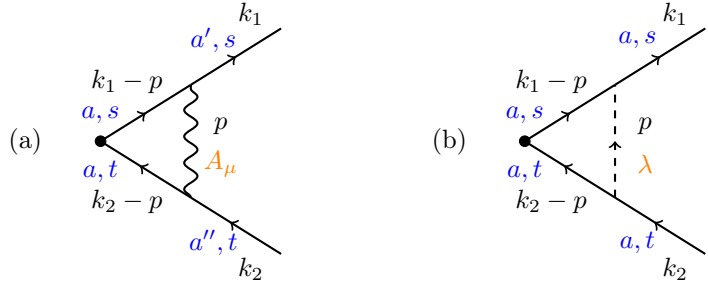

The following integral will contribute to panel (a):
\be
I_{A;2}=\int \frac{d^3p}{8\pi^3} G_B(k_1-p)G_B(k_2-p)(2k_1-p)_{\mu}(2k_2-p)_{\nu} \frac{1}{\Pi_{A,\mu\nu}(p)} \left(\delta_{\mu\nu}-\zeta \frac{p_{\mu}p_{\nu}}{p^2}\right)
\label{eq:PH_gauge_0}
\ee
To extract the divergence, we can simplify the calculation by choosing $k_1=k_2.$ The expression above then gives (more details are presented in appendix \ref{app:detail})
\be
I_{A;2} \rightarrow  - \frac{8}{(2N_b+N_f)\pi^2}(1-\zeta)\log k. 
\label{eq:IA2}
\ee 
Comparing with the tree level diagrams, there is an additional prefactor $3$ coming from the trace over gauge indices 
\be
\sum_j\sum_{a} \sigma^j_{a''a} \sigma^j_{aa'}= 3\delta_{a'a''}.
\ee
Another useful integral that contributes to panel (b) is, 
\be
I_{\lambda;2} = \i^2\int \frac{d^3p}{8\pi^3} G_B(k_1-p)G_B(-k_2+p)D_{\lambda}(p)
\rightarrow \frac{2}{N_b\pi^2}\log k, 
\label{eq:PH_lambda_0}
\ee
where we again have imposed $k_1=k_2$ and extracted the term proportional to $\log k$. Combining the contributions, we get
\be
\eta_{\text{vertex}}  =\frac{2}{N_b \pi^2}-\frac{8\cdot 3}{(2N_b+N_f)\pi^2} (1-\zeta).
\ee
The dimension of the quadratic boson term is thus, 
\be
\text{dim}[B^{\dagger}_{as} T^{\alpha}_{st} B_{at}]= (1+\eta_B)+\eta_{\text{vertex}}
=1+\frac{8}{3(2N_b+N_f)\pi^2}\left(\frac{N_f}{N_b}-22\right).
\ee 
At $N_f=N_b=2,$ we have above equal to $1-28/3\pi^2=0.054$. 
Our anomalous scaling dimension is 
\be
\eta_{B^2}=1+2\eta_B+2\eta_{\text{vertex}}=1+\frac{16}{3(2N_b+N_f)\pi^2}\left(\frac{N_f}{N_b}-22\right).
\ee

\subsubsection{Superconducting order parameter}

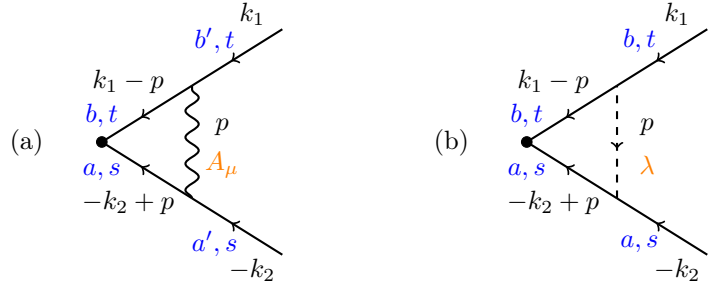
\begin{figure}[htbp]
\centering
\begin{tikzpicture}
\node at (-1,0) {(a)};
\filldraw (0,0) circle (2pt);
\draw[thick,-<-=0.5] (0,0)--(1.2,0.75);
\draw[thick,-<-=0.5] (1.2,0.75)--(2.4,1.5);
\draw[thick,-<-=0.5] (0,0)--(1.2,-0.75);
\draw[thick,-<-=0.5] (1.2,-0.75)--(2.4,-1.5);
\draw[thick, snake it] (1.2,-0.75)--(1.2,0.75);
\node at (0.35,0.8) {$k_1-p$};
\node at (0.,0.35) {\textcolor{blue}{$b, t$}};
\node at (0.35,-0.8) {$-k_2+p$};
\node at (0.,-0.4) {\textcolor{blue}{$a, s$}};
\node at (2,1.7) {$k_1$};
\node at (1.5,1.35) {\textcolor{blue}{$b',t$}};
\node at (2,-1.7) {$-k_2$};
\node at (1.5,-1.35) {\textcolor{blue}{$a',s$}};
\node at (1.6,0.2) {$p$};
\node at (1.6,-0.3) {$\textcolor{orange}{A_{\mu}}$};
\end{tikzpicture}
\quad\quad\quad\quad\quad
\begin{tikzpicture}
\node at (-1,0) {(b)};
\filldraw (0,0) circle (2pt);
\draw[thick,-<-=0.5] (0,0)--(1.2,0.75);
\draw[thick,-<-=0.5] (1.2,0.75)--(2.4,1.5);
\draw[thick,-<-=0.5] (0,0)--(1.2,-0.75);
\draw[thick,-<-=0.5] (1.2,-0.75)--(2.4,-1.5);
\draw[thick, dashed,-<-=0.5] (1.2,-0.75)--(1.2,0.75);
\node at (0.35,0.8) {$k_1-p$};
\node at (0.,0.35) {\textcolor{blue}{$b, t$}};
\node at (0.35,-0.8) {$-k_2+p$};
\node at (0.,-0.4) {\textcolor{blue}{$a, s$}};
\node at (2,1.7) {$k_1$};
\node at (1.5,1.35) {\textcolor{blue}{$b,t$}};
\node at (2,-1.7) {$-k_2$};
\node at (1.5,-1.35) {\textcolor{blue}{$a,s$}};
\node at (1.6,0.2) {$p$};
\node at (1.6,-0.3) {$\textcolor{orange}{\lambda}$};
\end{tikzpicture}
\caption{Vertex corrections to the superconducting order parameter.}
\label{fig:SC_0}
\end{figure}

Next we work out the vertex corrections to get the 
\be
\text{dim}\left[B_{as} \varepsilon_{ab} \mathcal{J}_{st}  B_{bt} \right]=2\ \text{dim}[B]+\iota_{\text{vertex}}.
\ee
At $v=0$, the result is guaranteed by symmetry to be the same as that of the charge order computed in the previous section, but we still present it here for completeness. The relevant diagrams are shown in \ref{fig:SC_0}. We will again first compute the integrals and then take into account the indices. 

The useful integral in panel (a) is
\be
\begin{split}
I_{A;3} = & \int \frac{d^3p}{8\pi^3} G_B(k_1-p)G_B(-k_2+p)(2k_1-p)_{\mu}(-2k_2+p)_{\nu} \frac{1}{\Pi_{A,\mu\nu}(p)} \left(\delta_{\mu\nu}-\zeta \frac{p_{\mu}p_{\nu}}{p^2}\right)\\
\rightarrow\ & \frac{8}{(2N_b+N_f)\pi^2}(1-\zeta) \log k.
\end{split}
\ee
Notice the integral is different from that in \eqref{eq:PH_gauge_0} and the result has opposite sign. 
Compared with the tree level, we just have an additional factor $-3$ coming from the gauge indices 
\be
\sum_{a,b} \sigma^j_{a'a} \varepsilon_{ab} (\sigma^j)^T_{bb'} = - 3\ \varepsilon_{a'b'}.
\ee
The two minus signs therefore cancel each other and we have the same result as in the charge density wave case. 

The integral relevant to panel (b) turns out to be have the same result same as that in \eqref{eq:PH_lambda_0}, 
\be
I_{\lambda;3} = \mathrm{i}^2\int \frac{d^3p}{8\pi^3}G_B(k_1-p)G_B(-k_2+p)D_{\lambda}(p)\\
=I_{\lambda;2}, 
\label{eq:I_lambda_3}
\ee 
with no additional prefactors compared with the tree level result. Combining all contributions, we get
\be
\iota_{\text{vertex}} =\frac{2}{N_b \pi^2}-\frac{8\cdot 3}{(2N_b+N_f)\pi^2} (1-\zeta), 
\ee
which is not surprisingly the same as that found in \eqref{eq:etaB_0}. The dimension of the pairing term is then
\be
\text{dim}[B_{as}\varepsilon_{ab}\mathcal{J}_{st}B_{bt}]=1+\frac{8}{3(2N_b+N_f)\pi^2}\left(\frac{N_f}{N_b}-22\right).
\ee 
Our anomalous scaling dimension is again
\be
\iota_{B^2}=1+2\eta_B+2\iota_{\text{vertex}}=1+\frac{16}{3(2N_b+N_f)\pi^2}\left(\frac{N_f}{N_b}-22\right).
\ee

\subsubsection{Correlation length exponent}

We compute the correlation length exponent $\nu$ of the order parameters,
\be
\xi \propto (g-g_c)^{-\nu}\,
\ee
following the method of Ref.~\onlinecite{KaulSS08}. As the correlation length is gauge-invariant, the calculation can be performed in a fixed gauge and $\nu=\nu_B$. We will use the relation 
\be
\nu_B=\frac{\gamma_B}{2-\eta_B},
\label{eq:scale_relation}
\ee
where the anomalous scaling dimension of single boson $\eta_B$ has been computed in \eqref{eq:etaB_0}, and $\gamma_B$ is defined as
\be
G_B^{-1}(k=0)=(g-g_c)^{\gamma_B}. 
\ee
We will calculate $\gamma_B$ below. We start by defining a convenient parameter $\lambda_g$ to measure the deviation from the critical point, satisfying 
\be
\frac{1}{g_c}-\frac{1}{g} = \frac{\sqrt{\lambda_g}}{4\pi}\,.
\ee
To leading order, it is related to $\bar{\lambda}$ by
\be
\bar{\lambda}=\lambda_g + \frac{\Pi_{\lambda}(k=0,\bar{\lambda}=0)}{\Pi_{\lambda}(0,\bar{\lambda})}\Sigma (0,0)\,,
\ee
where $\Sigma$ is the boson self energy, and its second argument refers to the mass in the boson propagator.
The boson propagator can then be written as
\be
G_B^{-1}(0) =\bar{\lambda}-\Sigma (0, \bar{\lambda}) =\lambda_g - \left( \Sigma (0,\lambda_g)-\frac{\Pi_{\lambda}(0,0)}{\Pi_{\lambda}(0,\lambda_g)}\Sigma (0,0)\right)\,,
\label{eq:braket}
\ee
where second argument of $\Pi_\lambda$ also refers to the boson mass.
In the following we will evaluate the $\lambda_g \log \lambda_g$ divergence of the self-energy diagrams appearing in \eqref{eq:braket}. 
\begin{figure}[htbp]
\centering
\begin{tikzpicture}[scale=0.7]
\node at (-5,2) {(a)};
\draw[thick,->-=0.5](-4,0) -- (-2,0);
\draw[thick,->-=0.5](-2,0) -- (2,0);
\draw[thick,->-=0.5](2,0) -- (4,0);
\draw[thick,snake it] (-2,-0.5) .. controls +(-0.4,3) and +(0.4,3) .. (2,-0.5);
\filldraw[fill=white,draw=white] (-2.5,-0.03)--(-1.5,-0.03)--(-1.5,-0.5)--(-2.5,-0.5);
\filldraw[fill=white,draw=white] (2.5,-0.03)--(1.5,-0.03)--(1.5,-0.5)--(2.5,-0.5);
\node at (-3,-0.4) {$k$};
\node at (3,-0.4) {$k$};
\node at (0,-0.4) {$k+p$};
\node at (1,0.8) {\textcolor{orange}{$A_{\mu}$}};
\node at (0,1.3) {$p$};
\end{tikzpicture}
\quad
\begin{tikzpicture}[scale=0.7]
\node at (-5,2) {(b)};
\draw[thick,->-=0.5](-4,0) -- (0,0);
\draw[thick,->-=0.5](0,0) -- (4,0);
\draw[thick, snake it] (0,-0) .. controls (-0.4,0.5) and (-1.4,2.2) ..(0,2.4).. controls (1.4,2.2) and (0.5,0.5)  .. (0,0);
\filldraw[fill=black]  (-0.05,0) circle (2pt);
\node at (-3,-0.4) {$k$};
\node at (3,-0.4) {$k$};
\node at (0,1.4) {\textcolor{orange}{$A_{\mu}$}};
\node at (1,1.3) {$p$};
\end{tikzpicture}
\\
\begin{tikzpicture}[scale=0.7]
\node at (-5,2) {(c)};
\draw[thick,->-=0.5](-4,0) -- (-2,0);
\draw[thick,->-=0.5](-2,0) -- (2,0);
\draw[thick,->-=0.5](2,0) -- (4,0);
\draw[thick,dashed,-<-=0.5] (-2,-0.5) .. controls +(-0.4,3.5) and +(0.4,3.5) .. (2,-0.5);
\filldraw[fill=white,draw=white] (-2.5,-0.03)--(-1.5,-0.03)--(-1.5,-0.5)--(-2.5,-0.5);
\filldraw[fill=white,draw=white] (2.5,-0.03)--(1.5,-0.03)--(1.5,-0.5)--(2.5,-0.5);
\node at (-3,-0.4) {$k$};
\node at (3,-0.4) {$k$};
\node at (0,-0.4) {$k+p$};
\node at (1.3,0.8) {\textcolor{orange}{$\lambda$}};
\node at (0,1.65) {$p$};
\end{tikzpicture}
\quad
\begin{tikzpicture}[scale=0.7]
\node at (-5,2) {(d)};
\draw[thick,->-=0.5](-4,0) -- (0,0);
\draw[thick,->-=0.5](0,0) -- (4,0);
\draw[thick,dashed] (0,0)--(0,0.8);
\draw[thick,-<-=0.3] (0,1.6) circle (0.8);
\draw[thick, snake it] (-0.8,1.6) -- (0.83,1.6);
\node at (-3,-0.4) {$k$};
\node at (3,-0.4) {$k$};
\node at (0.4,0.4) {$0$};
\node at (0,2.) {$q$};
\node at (1,1.1) {$p$};
\end{tikzpicture}
\\~\\
\begin{tikzpicture}[scale=0.7]
\node at (-5,2) {(e)};
\draw[thick,->-=0.5](-4,0) -- (0,0);
\draw[thick,->-=0.5](0,0) -- (4,0);
\draw[thick,dashed] (0,0)--(0,0.8);
\draw[thick,-<-=0.3] (0,1.6) circle (0.8);
\draw[thick, dashed] (-0.8,1.6) -- (0.8,1.6);
\node at (-3,-0.4) {$k$};
\node at (3,-0.4) {$k$};
\node at (0.4,0.4) {$0$};
\node at (0,2.) {$q$};
\node at (1,1.1) {$p$};
\end{tikzpicture}
\quad
\begin{tikzpicture}[scale=0.7]
\node at (-5,2) {(f)};
\draw[thick,->-=0.5](-4,0) -- (0,0);
\draw[thick,->-=0.5](0,0) -- (4,0);
\draw[thick,dashed] (0,0)--(0,0.8);
\draw[thick,-<-=0.3] (0,1.6) circle (0.8);
\draw[thick, snake it] (-0.8,1.63) .. controls (1,0.5) and (1,2.5)  ..(-0.8,1.6);
\filldraw[fill=black]  (-0.8,1.6) circle (2pt);
\node at (-3,-0.4) {$k$};
\node at (3,-0.4) {$k$};
\node at (0.4,0.4) {$0$};
\node at (0,1.5) {$q$};
\node at (1,1.1) {$p$};
\end{tikzpicture}
\caption{Feymann diagrams that contribute to the \eqref{eq:braket}. Gauge and flavor indices are supressed.}
\label{fig:self_energy}
\end{figure}
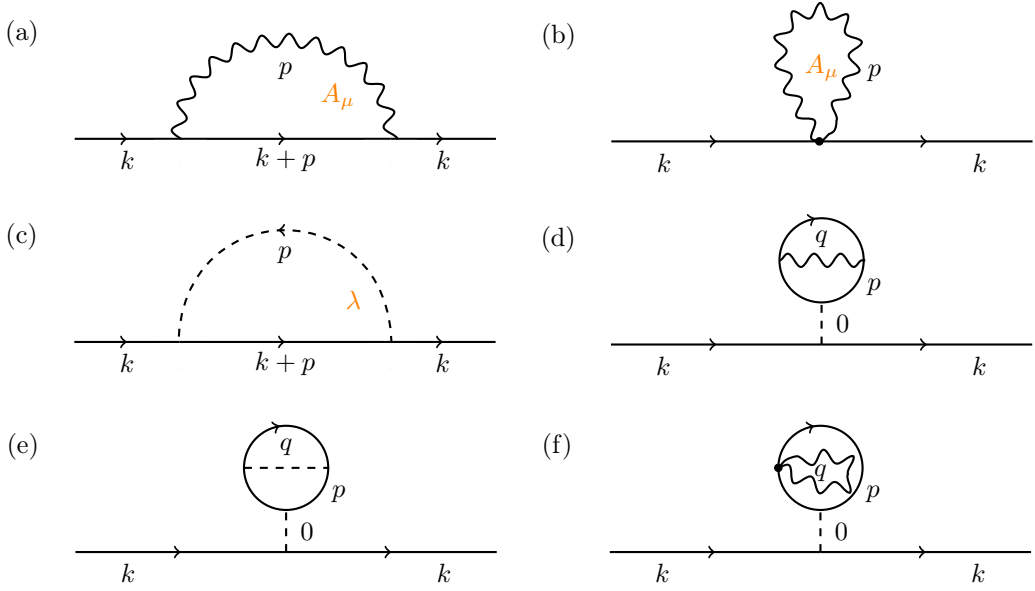
The relevant diagrams are shown in fig. \ref{fig:self_energy}, we list their contributions below:
\be
\begin{split}
& \Sigma^{(a)}= 3I_{A;1},\\
& \Sigma^{(b)}= 3\sum_{\mu,\nu}\int \frac{d^3p}{8\pi^3}\left(\delta_{\mu,\nu}-\zeta\frac{p_{\mu}p_{\nu}}{p^2}\right) \frac{1}{\Pi_A (p)}, \\
& \Sigma^{(c)}= I_{\lambda;1},\\
& \Sigma^{(d)}=  \frac{3\mathrm{i}^2}{\Pi_{\lambda}(0,\bar{\lambda})}   \int \frac{d^3p}{8\pi^3} I_{A;1}(p)\big(G_B (p)\big)^2\\
& \Sigma^{(e)}=  \frac{\mathrm{i}^2}{\Pi_{\lambda}(0,\bar{\lambda})} \int \frac{d^3p}{8\pi^3} I_{\lambda;1}(p)\big(G_B (p)\big)^2 \\
& \Sigma^{(f)}= \frac{\mathrm{i}^2}{\Pi_{\lambda}(0,\bar{\lambda})} \Sigma^{(b)}\int \frac{d^3p}{8\pi^3}\big(G_B (p)\big)^2 = -\Sigma^{(b)}.\\
\end{split}
\ee
where $I_{A;1}$ and $I_{\lambda;1}$ have been defined in the first equalities of the equations \eqref{eq:I_A_1} and \eqref{eq:I_lambda_1}, respectively. Since $(b)$ and $(f)$ cancel each other, we just need to extract the divergence in $(a)(c)(d)(e)$. The gauge field contributions $\Sigma^{(a)}+\Sigma^{(d)}$ give: 
\be
\Sigma^{(a)}+\Sigma^{(d)}\rightarrow -\frac{12}{\pi^2}\left(\frac{7N_f-18N_b}{(2N_b+N_f)^2} + \frac{\zeta}{2N_b+N_f}\right) \lambda_g \log \lambda_g.
\ee
The remaining term to be evaluated is $\Sigma^{(c)}+\Sigma^{(e)}$,
\be
\Sigma^{(c)}+\Sigma^{(e)}\rightarrow \frac{3}{\pi^2 N_b} \lambda_g \log \lambda_g.
\ee
Since the integrals are in parallel to those discussed in ref. \cite{KaulSS08}, we will omit the details here. Combining the two equations above, we have the total coefficient $\alpha$ in front of the $\lambda_g\log\lambda_g$ divergence as
\be
\alpha=\frac{3}{N_b \pi^2}-\frac{12}{\pi^2} \left(\frac{7N_f-18N_b}{(N_f+2N_b)^2}+\frac{\zeta}{N_f+2N_b}\right).
\ee 
Then we re-exponentiate the result and combine with equation \eqref{eq:braket} to get
\be
G^{-1}(0)=\lambda_g \left(1 - \alpha \log \frac{\lambda_g}{\Lambda^2} \right)\approx |g-g_c|^{2(1-\alpha)},
\ee
such that 
\be
\gamma_B=2-2\alpha=2-\frac{6}{N_b\pi^2}+\frac{24}{\pi^2} \left(\frac{7N_f-18N_b}{(N_f+2N_b)^2}+\frac{\zeta}{N_f+2N_b}\right).
\ee
Using the scaling relation \eqref{eq:scale_relation}, we get
\be
\nu_B\approx \frac{\gamma_B}{2} \left(1+\frac{\eta_B}{2}\right)\approx 1-\frac{8}{3N_b \pi^2}-\frac{20}{(2N_b+N_f)\pi^2}+\frac{12}{\pi^2}\frac{7N_f-18N_b}{(2N_b+N_f)^2}
\ee
where we have kept the leading terms. At $N_b=N_f=2,$ this gives $\nu_B=-0.216.$

\subsubsection{Dressed fermion field}

Only the gauge field contributes to the correction. The relevant integral is
\be
I_{A;\psi} 
=\int \frac{d^3q}{8\pi^3}\ \gamma_{\mu} G_{\psi}\gamma_{\nu}(k+q)D_{\mu\nu}(-q) 
\rightarrow \frac{8}{(N_f+N_b)\pi^2}\left(\frac{1}{3}-\zeta\right)\slashed{k}\log k.
\ee
Tracing over the gauge degrees of freedom, as in the boson case we just get a factor of three:
\be
\sum_{b,b'} (2\delta_{ab'}-\delta_{ab}\delta_{bb'})=4-1=3.
\ee
So the anomalous dimension for dressed fermion propagator is
\be
\eta_{\psi}=\frac{3\cdot 8}{(2N_b+N_f)\pi^2}\left(\frac{1}{3}-\zeta\right),
\ee
and the gauge-dependent fermion scaling dimension is
\be
\text{dim}[\psi]=1+\frac{\eta_{\psi}}{2}=1+\frac{3\cdot 4}{(2N_b+N_f)\pi^2}\left(\frac{1}{3}-\zeta\right).
\ee

\subsubsection{Boson-fermion composite}
The physical electron is a composite of bosonic chargon and fermionic spinon. We are interested in the scaling dimension of the electron operator at the four nodal points $\bm{k} = (\pm \pi / 2, \pm \pi / 2)$, whose precise representation in terms of the low-energy chargons and spinons is given in Eq.~\ref{cBpsi}. These are linear combinations of the operators $\sum_a B_{as}^* \psi_{at\alpha}$ and $\sum_{ab} \varepsilon_{ab} B_{as} \psi_{at\alpha}$, where $\alpha$ labels the spinor component. As we show below, the details of this linear combination are not essential as each of these terms are independently renormalized in the same manner.

\begin{figure}
\centering
\begin{tikzpicture}
\node at (-1,0) {(a)};
\draw[thick,->-=0.5] (0,0)--(1.2,0.75);
\draw[thick,->-=0.5] (1.2,0.75)--(2.4,1.5);
\draw[double,-<-=0.5] (0,0)--(1.2,-0.75);
\draw[double,-<-=0.5] (1.2,-0.75)--(2.4,-1.5);
\draw[thick, snake it] (1.2,-0.75)--(1.2,0.75);
\filldraw (0,0) circle (2pt);
\node at (0.5,0.7) {\textcolor{blue}{$a, t$}};
\node at (0.5,-0.7) {\textcolor{blue}{$a, s$}};
\node at (1.5,1.35) {\textcolor{blue}{$a',t$}};
\node at (1.5,-1.35) {\textcolor{blue}{$a'',s$}};
\node at (1.6,-0.1) {$\textcolor{orange}{A_{\mu}}$};
\end{tikzpicture}
\quad \quad \quad \quad
\begin{tikzpicture}
\node at (-1,0) {(a)};
\draw[thick,-<-=0.5] (0,0)--(1.2,0.75);
\draw[thick,-<-=0.5] (1.2,0.75)--(2.4,1.5);
\draw[double,-<-=0.5] (0,0)--(1.2,-0.75);
\draw[double,-<-=0.5] (1.2,-0.75)--(2.4,-1.5);
\draw[thick, snake it] (1.2,-0.75)--(1.2,0.75);
\filldraw (0,0) circle (2pt);
\node at (0.5,0.7) {\textcolor{blue}{$b, t$}};
\node at (0.5,-0.7) {\textcolor{blue}{$a, s$}};
\node at (1.5,1.35) {\textcolor{blue}{$b',t$}};
\node at (1.5,-1.35) {\textcolor{blue}{$a',s$}};
\node at (1.6,-0.1) {$\textcolor{orange}{A_{\mu}}$};
\end{tikzpicture}
\caption{Vertex corrections for the $B^{\dagger}\psi$ (left) and $B^T\psi$ (right) operators.}
\label{fig:electron}
\end{figure}
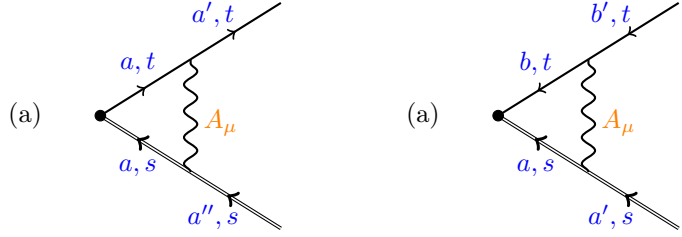

We first consider scaling corrections to the operator $\sum_a B_{as}^* \psi_{at}$. There exists a one-loop vertex correction by the gauge field shown in Fig. \ref{fig:electron}, leading to 
\be
\text{dim}[B^{\dagger}\psi]  =\text{dim}[B]+\text{dim}[\psi]+\eta_{B^{\dagger}\psi}
= \frac{3}{2} + \frac{1}{3(2N_b+N_f)\pi^2}\left(\frac{N_f}{N_b}-118\right). 
\label{eq:composite_0}
\ee
At $N_b=N_f=2$, this gives $\frac{3}{2}-\frac{13}{2\pi^2}=0.84.$
Importantly, this vertex correction in \ref{fig:electron} is unaffected by the presence of $\gamma$ matrices, so operators of the form $(1 \pm i \gamma^x) \sum_a B_{as}^* \psi_{at}$, which project to an individual spinor component, receive the same scaling dimension correction.

Another gauge-invariant choice is $\sum_{a,b} B_{as} \varepsilon_{ab} \psi_{bt}$, which gives the same contribution as in \eqref{eq:composite_0}. One can easily check that further adding $\gamma$ matrices acting in the spinor space of fermions doesn't change the result either. 

The scaling dimension of the quasiparticle residue of the electron Green's function, $Z$, is given by
\begin{align}
    \text{dim}[Z] = 2(\text{dim}[B^{\dagger}\psi] - 1)\,.
\end{align}

\subsubsection{Fermion bilinear}
The N\'eel and VBS correlation functions can be expressed in terms of spinon bilinear operators, and corrections to the scaling dimension of these operators can be calculated by analyzing the renormalization of these composite operators. In the absence of chargon fluctuations, corrections due to the $\text{SU}(2)$ gauge field have previously been computed~\cite{RanWen06}. To leading order, the only consequence of charge fluctuations is to modify the prefactor in the effective gauge propagator. The anomalous exponent is hence
\begin{equation}
    \eta_{\text{N\'eel, VBS}} = - \frac{16}{\pi^2 (N_f + 2 N_b)}
\end{equation}
where the results of~\cite{RanWen06} are recovered by setting $N_b = 0$.

\subsection{Finite $v$}
Next we consider nonzero $v$ in the Lagrangian \eqref{Lcomplete}. The matrix $M_b$ in \eqref{eq:F0} now becomes 
\be
\begin{split}
& \mathcal{G}^{-1}_b=\left(\begin{matrix}  \mathcal{G}^{-1}_A & \mathcal{G}^{-1}_B \\ \mathcal{G}^{-1}_C & \mathcal{G}^{-1}_D \end{matrix}
\right),\quad \mathcal{G}^{-1}_B=-2\Delta (k-k')\mathbbm{1},\quad \mathcal{G}^{-1}_C=-2\Delta^* (k'-k)\mathbbm{1},\\
& \mathcal{G}^{-1}_{A/D}=\mathbbm{1}\left[\delta_{kk'}k^2+\mathrm{i}\lambda (k-k')+\int \frac{d^3q}{(2\pi)^3}A_j(q)A_j(k-k'-q)\right]
\pm\sigma^j \left[(k+k')_{\mu}A_j^{\mu}(k-k')\right]
\end{split}
\ee
Notice that the matrix has indices $\{k,s,a;k',s',a'\}$, where $a, a'$ is the gauge index and $s, s'$ labels boson flavor. Trace performed over $k$ is simply a momentum integration. 
Further expanding the matrix log to second order, we obtain
\be
\begin{split}
\tr \ln \mathcal{G}_b^{-1} G_B= & N_b \int \frac{d^3p}{(2\pi)^3} \frac{d^3q}{(2\pi)^3}  \Bigg\{  G(q) \delta_{p0}  \int \frac{d^3 p'}{(2\pi)^3} A_{j\mu}(p')A_j^{\mu}(-p')\\
& -\frac{1}{2} G(q)G(q-p) \bigg[
- \lambda(p)\lambda(-p)+4\Delta(p)\Delta^*(p)+  \sum_j (2q-p)_{\mu} A_j^{\mu}(p)  (2q-p)_{\nu} A_j^{\nu}(-p) \bigg]\Bigg\}.\\
\end{split}
\ee
The fermionic sector is the same as before. Plugging in the integrals \eqref{eq:integrals}, we arrive at
\be
\begin{split}
{\mathcal{F}^{(1)}}= & \frac{1}{2}\int \frac{d^3p}{(2\pi)^3} \Bigg\{
\Pi_{\lambda} (p) \bigg[
 \lambda(p)\lambda(-p)-4\Delta(p)\Delta^*(p)\bigg]
\\
& +A_{j\mu}(p)\left(\delta_{\mu\nu}-\frac{p_{\mu}p_{\nu}}{p^2}\right)\Pi_A (p) A_j^{\nu}(-p)\Bigg\}+N_b\left(\frac{\lambda^2}{2u}-\frac{\lambda}{g}\right),
\end{split}
\ee
where the kernels are the same as before in \eqref{eq:kernels}. 
In addition to the gauge and $\lambda$ field propagators in \eqref{eq:propagators}, we now also have the $\Delta$ propagator 
\be
D_{\Delta}=-\frac{1}{4}D_{\lambda}=-\frac{1}{4\Pi_{\lambda}}.
\ee

\subsubsection{Dressed boson propagator}
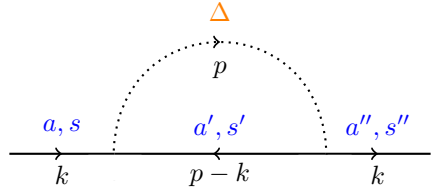
\begin{figure}[htbp]
\begin{tikzpicture}[scale=0.7]
\draw[thick,->-=0.5](-4,0) -- (-2,0);
\draw[thick,-<-=0.5](-2,0) -- (2,0);
\draw[thick,->-=0.5](2,0) -- (4,0);
\draw[thick,dotted,->-=0.5] (-2,-0.5) .. controls +(-0.4,3.5) and +(0.4,3.5) .. (2,-0.5);
\filldraw[fill=white,draw=white] (-2.5,-0.03)--(-1.5,-0.03)--(-1.5,-0.5)--(-2.5,-0.5);
\filldraw[fill=white,draw=white] (2.5,-0.03)--(1.5,-0.03)--(1.5,-0.5)--(2.5,-0.5);
\node at (-3,-0.4) {$k$};
\node at (3,-0.4) {$k$};
\node at (0,-0.4) {$p-k$};
\node at (-3,+0.5) {$\textcolor{blue}{a, s}$};
\node at (3,+0.5) {$\textcolor{blue}{a'', s''}$};
\node at (0,+0.5) {$\textcolor{blue}{a', s'}$};
\node at (0,1.6) {$p$};
\node at (0,2.7) {\textcolor{orange}{$\Delta$}};
\end{tikzpicture}
\caption{Additional correction due to $\Delta$ field, compared with fig.\ref{fig:single_B_0}.}
\label{fig:single_B}
\end{figure}

Now we have an additional diagram in fig. \ref{fig:single_B}.  
The relevant integral is simply $-\frac{1}{4} I_{\lambda;1}$ where $I_{\lambda;1}$ was computed in in \eqref{eq:I_lambda_1}. 
The trace over indices are computed as  
\be
\sum_{a'}\varepsilon_{aa'}
 \varepsilon_{a'a''} \sum_{s'} \mathcal{J}_{ss'} \mathcal{J}_{s's''}=\delta_{aa''}\delta_{ss''}, 
\ee
which is no different from the tree level diagram. Summing everything up, we now have
\be
\eta_{B}
=\frac{1}{2N_b \pi^2}-\frac{12}{(2N_b+N_f)\pi^2} \left(\frac{10}{3}+2\zeta \right).
\label{eq:etaB}
\ee
Dimension of single boson can be computed from 
$
\text{dim}[B]=(1+\eta_B)/2. 
$

\subsubsection{Density wave order parameter}
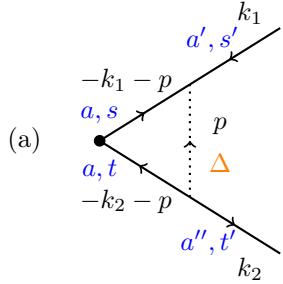
\begin{figure}[htbp]
\centering 
\begin{tikzpicture}
\node at (-1,0) {(a)};
\filldraw (0,0) circle (2pt);
\draw[thick,->-=0.5] (0,0)--(1.2,0.75);
\draw[thick,-<-=0.5] (1.2,0.75)--(2.4,1.5);
\draw[thick,-<-=0.5] (0,0)--(1.2,-0.75);
\draw[thick,->-=0.5] (1.2,-0.75)--(2.4,-1.5);
\draw[thick, dotted, ->-=0.5] (1.2,-0.75)--(1.2,0.75);
\node at (0.35,0.8) {$-k_1-p$};
\node at (0.,0.35) {\textcolor{blue}{$a, s$}};
\node at (0.35,-0.8) {$-k_2-p$};
\node at (0.,-0.4) {\textcolor{blue}{$a, t$}};
\node at (2,1.7) {$k_1$};
\node at (1.5,1.35) {\textcolor{blue}{$a',s'$}};
\node at (2,-1.7) {$k_2$};
\node at (1.45,-1.35) {\textcolor{blue}{$a'',t'$}};
\node at (1.6,0.2) {$p$};
\node at (1.6,-0.3) {$\textcolor{orange}{\Delta}$};
\end{tikzpicture}
\caption{Additional diagram contributing to the vertex correction of the density wave scaling dimension at leading order, compared with fig. \ref{fig:DW_0}. }
\label{fig:DW}
\end{figure}

The additional diagram compared with the $v=0$ case is shown in figure \ref{fig:DW}.
The relevant integral is $-\frac{1}{4}I_{\lambda;2}$ computed in \eqref{eq:IA2}.  Next we take care of the indices, 
\be
(\sum_{a}\varepsilon_{a'a}\varepsilon_{aa''})\left[\sum_{s,t} \mathcal{J}_{t't}T^{\alpha}_{ts} \mathcal{J}_{ss'}  \right]
= -\delta_{a',a''}
\left[-\sum_{s,t} (T^{\alpha})^T_{t't} \mathcal{J}_{ts}\mathcal{J}_{ss'}   \right]= -\delta_{a',a''}T^{\alpha}_{s't'}.
\ee
In the first equality we have used \eqref{eq:constraints}. Note that there is an additional sign compared with the tree level result. 
Combining the all the contributions from vertex corrections, we get
\be
\eta_{\text{vertex}} 
= \frac{5}{2N_b \pi^2}-\frac{24}{(2N_b+N_f)\pi^2} (1-\zeta).
\ee
The dimension of the quadratic boson term is thus, 
\be
\text{dim}[B^{\dagger}_{as} T^{\alpha}_{st} B_{at}]= (1+\eta_B)+\eta_{\text{vertex}}
=1+\frac{1}{(2N_b+N_f)\pi^2}\left(3\frac{N_f}{N_b}-58\right).
\ee
Taking $N_f=N_b=2$, the dimension is $1-55/6\pi^2=0.07.$ Our anomalous scaling dimension is 
\be
\eta_{B^2}=1+\frac{2}{(2N_b+N_f)\pi^2}\left(3\frac{N_f}{N_b}-58\right).
\ee

\subsubsection{Superconducting order parameter}

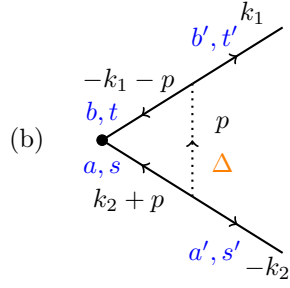
\begin{figure}[htbp]
\centering
\begin{tikzpicture}
\node at (-1,0) {(b)};
\filldraw (0,0) circle (2pt);
\draw[thick,-<-=0.5] (0,0)--(1.2,0.75);
\draw[thick,->-=0.5] (1.2,0.75)--(2.4,1.5);
\draw[thick,-<-=0.5] (0,0)--(1.2,-0.75);
\draw[thick,->-=0.5] (1.2,-0.75)--(2.4,-1.5);
\draw[thick, dotted,->-=0.5] (1.2,-0.75)--(1.2,0.75);
\node at (0.35,0.8) {$-k_1-p$};
\node at (0.,0.35) {\textcolor{blue}{$b, t$}};
\node at (0.35,-0.8) {$k_2+p$};
\node at (0.,-0.4) {\textcolor{blue}{$a, s$}};
\node at (2,1.7) {$k_1$};
\node at (1.5,1.35) {\textcolor{blue}{$b',t'$}};
\node at (2.2,-1.7) {$-k_2$};
\node at (1.5,-1.5) {\textcolor{blue}{$a',s'$}};
\node at (1.6,0.2) {$p$};
\node at (1.6,-0.3) {$\textcolor{orange}{\Delta}$};
\end{tikzpicture}
\caption{Additional contribution to the SC vertex correction, compared with fig. \ref{fig:SC_0}.}
\label{fig:SC}
\end{figure}

The additional diagram is figure \ref{fig:SC}. 
The relevant integral is the same as $-\frac{1}{4}I_{\lambda;2}$. Now we look at the indices,
\be
(\sum_{a,b}\varepsilon_{a'a}\varepsilon_{ab}\varepsilon_{bb'})(\sum_{s,t} \mathcal{J}_{s's}\mathcal{J}_{st}\mathcal{J}_{tt'})=\varepsilon_{a'b'}\mathcal{J}_{s't'},\ee
so again no additional prefactor is present compared with the tree level result. Combining all contributions, we get
\be
\iota_{\text{vertex}} 
= \frac{3}{2N_b \pi^2}-\frac{24}{(2N_b+N_f)\pi^2} (1-\zeta)
.
\ee
The dimension of this quadratic boson is thus
\be
\text{dim}\left[ B_{as} \varepsilon_{ab}  \mathcal{J}_{st} B_{bt} \right]= (1+\eta_B)+\iota_{\text{vertex}}
=1+\frac{2}{(2N_b+N_f)\pi^2}\left(\frac{N_f}{N_b}-30\right),
\ee
Taking $N_f=N_b=2$, the dimension is $1-29/3\pi^2=0.02.$ 
Anomalous scaling dimension is then
\be
\iota_{B^2}=1+\frac{4}{(2N_b+N_f)\pi^2}\left(\frac{N_f}{N_b}-30\right). 
\ee

\subsubsection{Correlation length exponent}

In addition to the diagrams presented in figures \ref{fig:self_energy}, now we also have figure \ref{fig:self_energy1} due to the $\Delta$ field:
\begin{figure}[htbp]
\centering
\begin{tikzpicture}[scale=0.7]
\node at (-5,2) {(g)};
\draw[thick,->-=0.5](-4,0) -- (-2,0);
\draw[thick,-<-=0.5](-2,0) -- (2,0);
\draw[thick,->-=0.5](2,0) -- (4,0);
\draw[thick,dotted,->-=0.5] (-2,-0.5) .. controls +(-0.4,3.5) and +(0.4,3.5) .. (2,-0.5);
\filldraw[fill=white,draw=white] (-2.5,-0.03)--(-1.5,-0.03)--(-1.5,-0.5)--(-2.5,-0.5);
\filldraw[fill=white,draw=white] (2.5,-0.03)--(1.5,-0.03)--(1.5,-0.5)--(2.5,-0.5);
\node at (-3,-0.4) {$k$};
\node at (3,-0.4) {$k$};
\node at (0,-0.4) {$p-k$};
\node at (-3,+0.5) {$\textcolor{blue}{a, s}$};
\node at (3,+0.5) {$\textcolor{blue}{a'', s''}$};
\node at (0,+0.5) {$\textcolor{blue}{a', s'}$};
\node at (0,1.6) {$p$};
\node at (0,2.7) {\textcolor{orange}{$\Delta$}};
\end{tikzpicture}
\quad 
\quad
\begin{tikzpicture}[scale=0.7]
\node at (-5,2) {(h)};
\draw[thick,->-=0.5](-4,0) -- (0,0);
\draw[thick,->-=0.5](0,0) -- (4,0);
\draw[thick,dashed] (0,0)--(0,0.8);
\draw[thick,-<-=0.3] (0,1.6) circle (0.8);
\draw[thick, dotted] (-0.8,1.6) -- (0.8,1.6);
\node at (-3,-0.4) {$k$};
\node at (3,-0.4) {$k$};
\node at (-3,+0.5) {$\textcolor{blue}{a, s}$};
\node at (3,+0.5) {$\textcolor{blue}{a, s}$};
\node at (0.4,0.4) {$0$};
\node at (0,2.) {$q$};
\node at (1,1.) {$p$};
\node at (-0.4,0.4) {$\textcolor{orange}{\lambda}$};
\end{tikzpicture}
\caption{Feymann diagrams in addition to fig. \ref{fig:self_energy} that constribute to $G^{-1}(0)$. }
\label{fig:self_energy1}
\end{figure}
Notice that the diagrams such as replacing the dashed $\lambda$ line in diagram \ref{fig:self_energy}(d) by the dotted $\Delta$ line are not allowed, because the two external boson propagators need to have the same gauge and flavor indices. The self energies corresponding to the above two diagrams are 
\be
\Sigma^{(g)}= -\frac{1}{4}\Sigma^{(c)},\quad 
\Sigma^{(h)}= -\frac{1}{4}\Sigma^{(e)}.
\ee
Both relevant integrals have been evaluated before, leading to 
\be
\Sigma^{(g)}+\Sigma^{(h)} \rightarrow -\frac{3}{4\pi^2 N_b}\lambda_g \log \lambda_g.
\ee
Combining with the $v=0$ results, we now have the modified total coefficient
\be
\alpha=\frac{9}{4N_b\pi^2}-\frac{12}{\pi^2} \left(\frac{7N_f-18N_b}{(N_f+2N_b)^2}+\frac{\zeta}{N_f+2N_b}\right).
\ee
The correlation function exponent is then (we also need to use the modified anomalous scaling dimension of $B$ in \eqref{eq:etaB}):
\be
\nu_B\approx  1+\frac{1}{N_b \pi^2}-\frac{20}{(2N_b+N_f)\pi^2}+\frac{12}{\pi^2}\frac{7N_f-18N_b}{(2N_b+N_f)^2}.
\ee
At $N_b=N_f=2$, this gives $\nu_B=-0.03.$

\subsubsection{Boson-fermion Composite}

One gauge invariant combination is $\sum_a B^{\dagger}_{as}\psi_{at}$. In the expression
$
\text{dim}[B^{\dagger}\psi]  =\text{dim}[B]+\text{dim}[\psi]+\eta_{B^{\dagger}\psi}= \frac{3}{2}+\frac{\eta_B+\eta_{\psi}}{2}+\eta_{B^{\dagger}\psi},
$
the only change compared with \eqref{eq:composite_0} lies in dim$[B]$.
The result is thus
\be
\text{dim}[B^{\dagger}\psi]  = 
\frac{3}{2} + \frac{1}{4(2N_b+N_f)\pi^2}\left(\frac{N_f}{N_b}-158\right). 
\label{eq:composite}
\ee
Other gauge-invariant choices such as $\sum_{a,b} B_{as} \varepsilon_{ab} \psi_{bt}$, or with $\gamma$ matrices inserted give the same results as in \eqref{eq:composite}. At $N_f=N_b=2$, the above expression gives dim$[B^{\dagger}\psi]=0.837.$
\renewcommand{\arraystretch}{2.5}
\begin{table}
\centering
\begin{tabular}{ l| c | c | c }
 & $v=0$ & SC channel & DW channel \\
 \hline
dim[$B_{as}^{\dagger}T_{st}B_{at}]$\ & $1+\dfrac{8\left(c-22\right)}{3(2N_b+N_f)\pi^2}$ & $1+\dfrac{\left(3c-58\right)}{(2N_b+N_f)\pi^2}$ & $1-\dfrac{16\left(c+14\right)}{3(2N_b+N_f)\pi^2}$ \\
\hline
dim$[B_{as}\varepsilon_{ab}\mathcal{J}_{st}B_{bt}]$\ & $1+\dfrac{8\left(c-22\right)}{3(2N_b+N_f)\pi^2}$ & $1+\dfrac{2\left(c-30\right)}{(2N_b+N_f)\pi^2}$  & $1+\dfrac{4\left(5c-38\right)}{3(2N_b+N_f)\pi^2}$ \\
\hline
dim$[B^{\dagger}\psi]$\ & $\dfrac{3}{2} + \dfrac{\left(c-118\right)}{3(2N_b+N_f)\pi^2}$ & $\dfrac{3}{2} + \dfrac{\left(c-158\right)}{4(2N_b+N_f)\pi^2}$ & $\dfrac{3}{2} - \dfrac{2\left(c+62\right)}{3(2N_b+N_f)\pi^2}$ \\
\hline
\quad $\nu_B$\ & 
$\begin{aligned}[t]
1 & - \dfrac{8}{3 N_b\pi^2} - \dfrac{20}{(2 N_b + N_f)\pi^2} \\
& + \dfrac{12}{\pi^2}\dfrac{N_b(7 c - 18 )} {(2 N_b + N_f)^2}\\    
\end{aligned}$
& $\begin{aligned}[t]
1 & + \dfrac {1}{N_b\pi^2} - \dfrac {20}{(2 N_b + N_f)\pi^2} \\
& + \dfrac {12} {\pi^2}\dfrac {N_b(7 c - 18 )} {(2 N_b + N_f)^2}
\end{aligned}$ 
& $\begin{aligned}[t] 1 & + \dfrac{7}{3 N_b\pi^2} - \dfrac {20}{(2 N_b + N_f)\pi^2} \\
 & + \dfrac{12}{\pi^2}\dfrac{N_b(7 c - 18 )}{(2 N_b + N_f)^2} \end{aligned}$ \\
\end{tabular}
\caption{Summary of scaling dimensions at the multicritical point (second column), with the Lagrangians in \eqref{LBa} (third column) and \eqref{e2a} (last column), respectively. $c\equiv N_f/N_b$ is a constant.}
\label{tab:dimensions}
\end{table}
\renewcommand{\arraystretch}{1}

We summarize the calculations of scaling dimensions in Table \ref{tab:dimensions}. 

\section{Honeycomb lattice}
\label{sec:honeycomb}

The ground state of the large-$U$ Hubbard model on the honeycomb lattice at half-filling has long-range N\'eel order, as for the square lattice. Also as for the square lattice, adding frustrating interactions leads to a phase with VBS ({\it i.e.} Kekul\'e) order \cite{NRSS90,Fu11,Yao19}. But in contrast to the square lattice, at smaller $U$ the honeycomb lattice features a semi-metal phase with no broken symmetry, and an electronic dispersion with 2 massless Dirac fermion points in the Brillouin zone. 

In this section we extend the SU(2) gauge theory analysis to the honeycomb lattice. We find just the three phases noted above, with no additional superconducting or charge-ordered phases. This difference from the square lattice case can be traced to the fact that the bosonic chargons, $B$,  move in a background zero flux on the honeycomb lattice \cite{Hermele07}. 
Consequently, the $B$ dispersion has only a single minimum in the Brillouin zone, and the Higgs phase where $B$ is condensed breaks no symmetries and realizes the Dirac semi-metal.
We sketched a phase diagram for the honeycomb lattice SU(2) gauge theory in Fig.~\ref{fig:honeycomb}. 

The details of such a theory have previously been worked out in Ref.~\onlinecite{Hermele07}, but with the interpretation of the deconfined phase as being stable - our interpretation is that this phase is ultimately unstable to either N\'eel or VBS order. The low energy theory consists of $N_f = 2$ Dirac fermions with an emergent $\text{SO}(5)$ symmetry rotating between N\'eel and Kekul\'e VBS order. As there is only a single minima of the chargon disperion at $\bm{k} = (0, 0)$, the spinons are coupled to $N_b=1$ bosonic chargons, with the full symmetry of the low-energy action being $\text{SO}(5) \times \text{SU}(2)$, with the $\text{SU}(2)$ chargon symmetry corresponding to the pseudospin. An important point which is not explicitly discussed in Ref.~\onlinecite{Hermele07} is the possibility of symmetry-allowed quartic interactions between the chargons and spinons, which would be marginal at tree level. However, this is rather simple to rule out due to the fact that the chargon minima is at $\bm{k} = (0, 0)$, and hence transforms trivially under all the lattice symmetries (an exception are transformations which exchange the $A$ and $B$ sublattice, where the sublattice structure of the chargon eigenvalue causes the chargon to acquire a minus sign - this has no effect on chargon bilinears). As a result, symmetry-allowed chargon/spinon quartic interactions demand that the spinon bilinear component is independently allowed by symmetry, and one can easily verify that no such term exists.

The large-$N_f\,, N_b$ expansion proceeds identically to the one discussed previously in the paper, with the exception that the chargon sector does not contain any quartic interactions aside from a $B^4$ term (in other words, we take $v=0$). The results for the various scaling dimensions in Section~\ref{sec:1N} carry over to this scenario, although some of the chargon bilinears studied can only be defined for even $N_b$.

We note an interesting relation between the model of Ref.~\onlinecite{Sato:2022lxm} and the SU(2) gauge field theory with $N_f=2$ and $N_b=1$. The global symmetry of the quantum field theory of Section~\ref{sec:introqft} is SO(5)$_f$ in the fermionic sector for $N_f=2$, and USp(2)/$\mathbb{Z}_2$ in the bosonic sector for $N_b=1$.
Ref.~\onlinecite{Sato:2022lxm} considered a honeycomb lattice model in which quantum spin Hall, superconducting, and Dirac semi-metal phases meet at a multicritical point, and proposed a SO(5) Gross-Neveu-Yukawa field theory for the multicriticality. The GNY field theory has no gauge fields, and hence there is an additional SO(3)$\cong$USp(2)/$\mathbb{Z}_2$ global symmetry which acts on the Dirac fermions. So the global symmetries of our SU(2) gauge field theory at  $N_f=2$ and $N_b=1$ are identical to those of the SO(5) GNY theory. It remains an interesting open question whether these two theories are the same conformal field theory.

\section{Discussion}

The discovery of high temperature superconductivity in the cuprates sparked decades of theoretical work on quantum phases proximate to the familiar N\'eel ordered state of the $S=1/2$ square lattice antiferromagnet. Early work \cite{NRSS89} argued that the proximate insulator has valence bond solid (VBS) order. The nature of the N\'eel-VBS quantum transition has also been extensively studied \cite{senthil1,senthil2,DQCP3}, and recent fuzzy sphere investigations \cite{Zhou:2023qfi} have concluded that it is described by a `pseudo-critical' theory with an approximate conformal symmetry, and a nearly exact global SO(5) symmetry which rotates between the $3+2$ components of the N\'eel and VBS orders. One formulation of the pseudo-critical theory has a SU(2) gauge field coupled to $N_f=2$ fundamental massless Dirac fermions: we have used the fuzzy sphere results to conclude that this gauge theory confines in the infrared with either N\'eel or VBS order, and the N\'eel-VBS transition is weakly first order. The ordering is selected by terms which are formally irrelevant in the continuum theory, and we assume here that N\'eel order is selected. 

The present paper extends these investigations by allowing for charge fluctuations, while remaining at half-filling and preserving particle-hole symmetry. Following earlier work \cite{PNAS_pseudo}, we have shown that adding charge fluctuations to the SU(2) gauge theory leads naturally to a $d$-wave superconductor with nodal quasiparticles, and states with period-2 charge order. We can then consider quantum transitions between the N\'eel state and the $d$-wave superconductor, or between the N\'eel state and charge order. Such transitions are described by a direct extension of the SU(2) gauge theory with $N_f=2$ fundamental massless Dirac fermions---there are additional fundamental $N_b=2$ massless complex scalars.  Given the weakly broken conformal symmetry for $N_f = 2$, $N_b = 0$ \cite{Sandvik20,Zhao20,Zhou:2023qfi}, and the stability of conformal gauge theories at large $N_{f,b}$, it is very plausible that the $N_f =2$, $N_b=2$ case exhibits true deconfined criticality with an exact emergent conformal symmetry.

The $N_f=2$, $N_b=2$ quantum field theory studied in this paper is defined by the Lagrangian $\mathcal{L}_\psi + \mathcal{L}_B$ in (\ref{LF}) and (\ref{LB}). Here $r$ is the tuning parameter which takes the system from the N\'eel state (present when $r$ is large and positive and $B$ is not condensed) to the states allowed by charge fluctuations (with $d$-wave superconductivity or charge order). The coefficients of the quartic couplings $v_{1,2,3}$ in (\ref{LB}) select among the latter states. 

We studied two different large $N_{f,b}$ generalizations of this theory, defined by the extensions (\ref{e2}) and (\ref{e2aa}) in the bosonic sector. The phase diagrams of these theories at $N_b = \infty$ appear in Fig.~\ref{fig:1}. The $1/N_{f,b}$ expansions of the second-order quantum phase transitions are described in Section~\ref{sec:1N} and Appendix~\ref{app:alternative}. We computed the scaling dimensions of the gauge-invariant order parameters, which are composites of two fermions or two bosons, and the electron operators at momenta $(\pm \pi/2, \pm \pi/2)$, which are the composites of one fermion and one boson in (\ref{cBpsi}). Our results are summarized in Table~\ref{tab:dimensions}. The results are not expected to be accurate at $N_f=N_b=2$, when the $1/N_{f,b}$ corrections are quite large.

The scaling dimension of the electron operator determines a novel feature of the quantum transition out of the $d$-wave superconductor. The $d$-wave superconductor itself is conventional, and has 4 nodal points with gapless Bogoliubov quasiparticles. In BCS theory, such gapless quasiparticles are remnants of the Fermi surface of the parent metal, and so the electronic quasiparticle residue remains non-zero across the metal-superconducting transition. However, for the transition from the $d$-wave superconductor to the N\'eel state, there is no longer a simple relationship between the Bogoliubov quasiparticles and the Fermi surface excitations of a parent metal. Instead, the Bogliubov quasiparticles of the superconductor are connected to the spinons of the deconfined quantum critical point. 
As there are no gapless electronic excitations in the N\'eel state, and the electronic quasiparticle residue vanishes at the transition out of the $d$-waves superconductor with an exponent determined by the scaling dimension of the electron operator at the deconfined quantum critical point. 

A recent paper \cite{Christos23} has shown that a similar phenomenon can also happen in the {\it electron} doped case in a situation where the normal state has no Fermi surface crossing the zone diagonals: nevertheless, gapless nodal quasiparticles do appear in the proximate $d$-wave superconductor, in a region of the Brillouin zone which is gapped in the normal state. 
Furthermore, there are connections of this remarkable phenomenon to the recent photoemission observations of Ref.~\onlinecite{Shen_2023} on the electron doped cuprates.

Along the same lines, we believe the $d$-wave superconductor found in the quasi-one-dimensional numerical study of Ref.~\onlinecite{Jiang23}, by doping the spin liquid of the $J_1$-$J_2$ antiferromagnet, will have 4 nodal points in the two-dimensional limit.

Finally, we note the analysis of Section~\ref{sec:honeycomb}, where we applied the same line of thought to the N\'eel-VBS transition on the honeycomb lattice \cite{NRSS90,Fu11}. We found only a single additional phase upon including charge fluctuations: a Dirac semi-metal with no broken symmetries. 
All these phases (N\'eel, VBS (Kekul\'e), Dirac semi-metal) have been observed in experiments on monolayer graphene \cite{Young14,Sacepe22}.
It is interesting to speculate that the absence of a superconducting phase on the honeycomb lattice in our theory, in contrast to the square lattice, is the underlying reason for the low superconducting $T_c$'s observed in the graphene family of compounds. 
~\\
\\
{\bf Acknowledgements}
~\\

We thank Marc-Henri Julien \cite{Julien_2024} and Cyril Proust and Bill Atkinson \cite{Proust_2024} for sharing their results before publication, and for useful discussions.
We also thank Fakher Assaad, Bernhard Keimer, Steve Kivelson, Anders Sandvik, T.~Senthil, and Z.-X.~Shen for useful discussions.
This research was supported by the U.S. National Science Foundation grant No. DMR-2245246 and by the Simons Collaboration on Ultra-Quantum Matter which is a grant from the Simons Foundation (651440, S.S.).

 \appendix

\section{Alternative Large $N_b$ Limit}
\label{app:alternative}

This appendix considers an alternative large $N_b$ limit of the $N_b=2$ case of the action $\mathcal{L}_B$ in (\ref{e2}).
We use the identity (\ref{identity}) to write (\ref{e2}) at $N_b=2$ as
\begin{align}
\widetilde{\mathcal{L}}_B =  \left| \left( \partial_\mu - i A_\mu^\alpha \sigma^\alpha \right) B \right|^2  +   \frac{u}{2N_b} \, (|B_{as}|^2 - 1/g)^2 +
\frac{v}{N_b} \, \left( B^\ast_{as} \sigma^i_{st} B_{at} \right)^2 \,. \label{e2a}
\end{align}
(For simplicity, we have ignored a renormalization of the values of $u$ and $g$ arising from the l.h.s. of (\ref{identity}).) Note that $v$ now appears with the opposite sign in the last quartic term compared to (\ref{e2}). The form (\ref{e2a}) is not suitable for a large $N_b$ generalization because it has `flavor' Pauli matrices which will generalize to the $N_b^2-1$ generators of SU$(N_b)$. To over come this difficulty, we use the following $N_b=2$ identity to transfer the Pauli matrices from the flavor to the gauge indices
\begin{align}
\sigma^i_{ss'} \sigma^i_{tt'} B^\ast_{as} B^\ast_{bt} B_{bt'} B_{as'} = \sigma^j_{aa'} \sigma^j_{bb'} B^\ast_{as} B^\ast_{bt} B_{b't} B_{a's}\,. \label{e10}
\end{align}
Here the index $j=1,2,3$ labels the adjoint gauge SU(2) components; (\ref{e10}) can be established by applying the following identity to both sides:
\begin{align}
\sigma^i_{ss'} \sigma^i_{tt'} = -\delta_{ss'} \delta_{tt'} + 2 \delta_{st'} \delta_{ts'}\,.
\end{align}
Then we can write (\ref{e2a}) as 
\begin{align}
\widetilde{\mathcal{L}}_B =  \left| \left( \partial_\mu - i A_\mu^\alpha \sigma^\alpha \right) B \right|^2  +   \frac{u}{2N_b} \, (|B_{as}|^2 - N_b/g)^2 +
\frac{v}{N_b} \, \left( B^\ast_{as} \sigma^j_{aa'} B_{a's} \right)^2 \,, \label{e2aa}
\end{align}
and the flavor indices $s,t$ can be extended to range over general $N_b$ values. The theory $\widetilde{\mathcal{L}}_B$ in (\ref{e2aa}) has a SU$(N_b) \times$U(1) global symmetry, in contrast to the theory $\mathcal{L}_B$ in (\ref{e2}) with a 
USp$(N_b) \times$U(1) global symmetry. By construction, the two theories are the same at $N_b=2$, but are distinct for $N_b >2$.

We can now proceed with a large $N_b$ expansion of (\ref{e2aa}). We decouple the $v$ term in (\ref{e2aa}) by a real Higgs field $H_j$ which is an adjoint under gauge SU(2), but a singlet under flavor SU($N_b$). In this manner we obtain, in place of (\ref{CondensedSP}), the action 
\begin{equation}
    \widetilde{S}_{\text{eff.}}= \frac{N_b}{2} \text{Tr}\left[\text{ln}(\mathcal{G}^{-1})\right]+\frac{N_b\lambda^2}{2u}-\frac{N_b H_j^2}{4v}-i\lambda\frac{N_b}{g}+i\lambda (|B_{a1}|^2+|B_{a2}|^2)-H_j B^*_{a1}\sigma^j_{aa'}B_{a'1}-H_j B^*_{a2}\sigma^j_{aa'}B_{a'2}
\end{equation}

where
\begin{equation}
    \mathcal{G}^{-1}=\begin{pmatrix}
        i\lambda-(\partial_{\mu}+iA_{\mu}^j\sigma_j)^2 -H_j\sigma^j& 0\\ 0 & i\lambda-(\partial_{\mu}-iA_{\mu}^j\sigma_j^T)^2 +H_j\sigma^j
    \end{pmatrix}
\end{equation}
is a $4 \times 4$ matrix. AS in the main text, we assume $A_\mu^j = 0$ at the saddle point.

The saddle point equation for $\lambda$ is as before after interchanging $4|\Delta|^2$ for $H_j^2$:
\begin{equation}\label{LambdaSPAPP}
    \frac{i\lambda N_b}{u}+\frac{N_b}{g}-N_b\left(|B_1|^2+|B_2|^2\right)=\int\frac{d^3p}{(2\pi)^3}\left(i\lambda+p^2\right)\frac{2N_b}{(i\lambda+p^2)^2-H_j^2}\,,
\end{equation}
and that for $H_j$ is identical to the saddle point for $\Delta$ after after interchanging $4|\Delta|^2$ for $H_j^2$ and taking $v\rightarrow-v$:
\begin{equation}\label{DeltaSPAPP}
    -\frac{N_b}{2v}-\frac{N_b}{H_z}\left(B^*_{a,1}\sigma^j_{aa'}B_{a',1}+B^*_{a,2}\sigma^j_{aa'}B_{a',2}\right)=\int\frac{d^3p}{(2\pi)^3}\frac{2N_b}{(i\lambda+p^2)^2-H_j^2} \,.
\end{equation}
In working out the saddle point equation for $B$, we will assume that if $H_j$ is condensed, it aligns only with $H_z$ to simplify the saddle point equations. Under this assumption, we obtain the saddle point equation for $B$:
\begin{equation}
    i\lambda B_{1,1}=H_zB_{1,1} \qquad i\lambda B_{2,1}=-H_zB_{2,1} \qquad i\lambda B_{1,2}=H_zB_{1,2} \qquad i\lambda B_{2,2}=-H_zB_{2,2} 
\end{equation}

After integration, these saddle point equations for $H_z$ and $\overline{\lambda} \equiv i \lambda$  obtained from integrating (\ref{DeltaSPAPP}) and (\ref{LambdaSPAPP}) are:
\begin{equation}\label{HeisSPAPP}
    -\frac{1}{2v}-\frac{1}{H_z}\left(B^*_{a,1}\sigma^z_{aa'}B_{a',1}+B^*_{a,2}\sigma^z_{aa'}B_{a',2}\right)=\frac{1}{4\pi |H_z|}\left[\sqrt{\overline{\lambda}+|H_z|}-\sqrt{\overline{\lambda}-|H_z|}\right]
\end{equation}
\begin{equation}
    \frac{\overline{\lambda}}{u}+\frac{1}{g}-(|B_{1}|^2+|B_2|^2)=-\frac{1}{4\pi}\left[\sqrt{\overline{\lambda}+|H_z|}+\sqrt{\overline{\lambda}-|H_z|}-\frac{4\pi}{g_c}\right]\,,
\end{equation}
where $1/g_c=\Lambda/\pi^2$, with $\Lambda$ the momentum space cutoff. If we set $B$ to zero then we will find all the same saddle point solutions in the main text where $B=0$ if we exchange $v\rightarrow-v$ and $2|\Delta|\rightarrow H_z$. If we allow $B$, $H_z$, and $\lambda$ to all condense, we will find that the saddle point equations will again enforce $\lambda=|H_z|$ and the saddle point equation for $H_z$ can be rewritten as:
\begin{equation}
    -\frac{1}{2v}-\frac{1}{\overline{\lambda}}\left(|B_1|^2+|B_2|^2\right)=\frac{\sqrt{2}}{4\pi\sqrt{\overline{\lambda}}}
\end{equation}
This is again the same as our previous saddle point equation for $\Delta$ if we exchange $v$ with $-v$ and $2|\Delta|$ with $|H_z|=\lambda$. We also note the types of solutions we find when $B$ is condensed. An example solution which solves the saddle point equation for $B$ when $H_z$ is nonzero has:
\begin{equation}
    B_{1,1}\neq 0 \qquad B_{1,2}=0 \qquad B_{2,1/2}=0 
\end{equation}
Such a solution will condense the CDW$_x$ order parameter in \cite{Christos23}. We could also have chosen a different example solution for $B$:
\begin{equation}
    B_{1,1}=B_{1,2}\neq 0 \qquad B_{2,1,2}=0 
\end{equation}
which would result in condensing the CDW$_y$ order in \cite{Christos23}:

Finally we could have chosen a solution where only the $d$-density wave is condensed with:
\begin{equation}
    B_{1,1}\propto \i \qquad B_{1,2}\propto 1 \qquad B_{2,1/2}=0 
\end{equation}
A general solution will have different nonzero strengths for each of the above continuum order parameters. There is no solution allowed by the saddle point equations where the $d$-wave pairing continuum order parameter is also condensed. The phase diagram for this large $N_b$ limit is shown in Fig.~\ref{fig:1}b.

\subsection{Large-$N$ corrections for the alternative formulation}

The leading correction to the free energy is now
\be
\begin{split}
{\mathcal{F}^{(1)}}= & \frac{1}{2}\int \frac{d^3p}{(2\pi)^3} \Bigg\{
\Pi_{\lambda} (p) \bigg[
 \lambda(p)\lambda(-p)
 - H_j(p) H_j(-p)\bigg]
\\
& +A_{j\mu}(p)\left(\delta_{\mu\nu}-\frac{p_{\mu}p_{\nu}}{p^2}\right)\Pi_A (p) A_j^{\nu}(-p)\Bigg\}+N_b\left(\frac{\lambda^2}{2u}+\frac{H_j^2}{2w}
-\frac{\lambda}{g}\right),
\end{split}
\ee
We now have the propagator of the $H_j$ fields
\be
D_H=-D_{\lambda}=-\frac{1}{\Pi_{\lambda}}. 
\ee

\subsubsection{Dressed boson propagator}
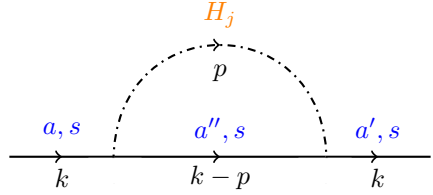
\begin{figure}[htbp]
\begin{tikzpicture}[scale=0.7]
\draw[thick,->-=0.5](-4,0) -- (-2,0);
\draw[thick,->-=0.5](-2,0) -- (2,0);
\draw[thick,->-=0.5](2,0) -- (4,0);
\draw[thick,dashdotted,->-=0.5] (-2,-0.5) .. controls +(-0.4,3.5) and +(0.4,3.5) .. (2,-0.5);
\filldraw[fill=white,draw=white] (-2.5,-0.03)--(-1.5,-0.03)--(-1.5,-0.5)--(-2.5,-0.5);
\filldraw[fill=white,draw=white] (2.5,-0.03)--(1.5,-0.03)--(1.5,-0.5)--(2.5,-0.5);
\node at (-3,-0.4) {$k$};
\node at (3,-0.4) {$k$};
\node at (0,-0.4) {$k-p$};
\node at (-3,+0.5) {$\textcolor{blue}{a, s}$};
\node at (3,+0.5) {$\textcolor{blue}{a', s}$};
\node at (0,+0.5) {$\textcolor{blue}{a'', s}$};
\node at (0,1.6) {$p$};
\node at (0,2.7) {\textcolor{orange}{$H_j$}};
\end{tikzpicture}
\caption{Additional correction due to $\Delta$ field, compared with fig.\ref{fig:single_B_0}.}
\label{fig:single_B1}
\end{figure}

Now we have an additional diagram in fig. \ref{fig:single_B1}.  

The relevant integral is simply $-I_{\lambda;1}$ where $I_{\lambda;1}$ was computed in in \eqref{eq:I_lambda_1}. 
The trace over indices simply gives an additional factor of three, such that 
\be
\begin{split}
& \eta_{B}=\frac{2}{3N_b \pi^2}(1-3)-\frac{4\cdot 3}{(2N_b+N_f)\pi^2} \left(\frac{10}{3}+2\zeta \right)\\
& =-\frac{4}{3N_b \pi^2}-\frac{12}{(2N_b+N_f)\pi^2} \left(\frac{10}{3}+2\zeta \right).\\
\end{split}
\label{eq:etaB2}
\ee
Dimension of single boson can be computed from 
$
\text{dim}[B]=(1+\eta_B)/2. 
$

\subsubsection{Density wave order parameter}
\begin{figure}[htbp]
\centering
\begin{tikzpicture}
\node at (-1,0) {(a)};
\filldraw (0,0) circle (2pt);
\draw[thick,->-=0.5] (0,0)--(1.2,0.75);
\draw[thick,->-=0.5] (1.2,0.75)--(2.4,1.5);
\draw[thick,-<-=0.5] (0,0)--(1.2,-0.75);
\draw[thick,-<-=0.5] (1.2,-0.75)--(2.4,-1.5);
\draw[thick, dashdotted, ->-=0.5] (1.2,-0.75)--(1.2,0.75);
\node at (0.35,0.8) {$k_1-p$};
\node at (0.,0.35) {\textcolor{blue}{$a, s$}};
\node at (0.35,-0.8) {$k_2-p$};
\node at (0.,-0.4) {\textcolor{blue}{$a, t$}};
\node at (2,1.7) {$k_1$};
\node at (1.5,1.35) {\textcolor{blue}{$a',s$}};
\node at (2,-1.7) {$k_2$};
\node at (1.45,-1.35) {\textcolor{blue}{$a'',t$}};
\node at (1.6,0.2) {$p$};
\node at (1.6,-0.3) {$\textcolor{orange}{H_j}$};
\end{tikzpicture}
\caption{Additional diagram contributing to the vertex correction of the density wave scaling dimension at leading order, compared with fig. \ref{fig:DW_0}. }
\label{fig:DW1}
\end{figure}
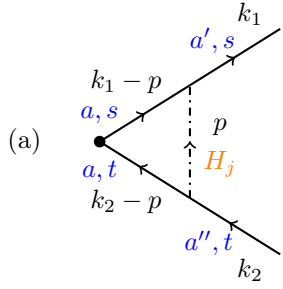

The additional relevant diagram is shown in figure \ref{fig:DW1}.
The relevant integral is $-I_{\lambda;2}$ computed in \eqref{eq:IA2}. The indices gives a factor of three, leading to 
\be
\begin{split}
\eta_{\text{vertex}} & =\frac{2}{N_b \pi^2}(1-3)-\frac{8\cdot 3}{(2N_b+N_f)\pi^2} (1-\zeta)\\
& = -\frac{4}{N_b \pi^2}-\frac{24}{(2N_b+N_f)\pi^2} (1-\zeta)
.
\end{split}
\ee
The dimension of the quadratic boson term is thus, 
\be
\text{dim}[B^{\dagger}_{as} T^{\alpha}_{st} B_{at}]= (1+\eta_B)+\eta_{\text{vertex}}
=1-\frac{16}{3(2N_b+N_f)\pi^2}\left(\frac{N_f}{N_b}+14\right).
\ee
Taking $N_f=N_b=2$, the dimension is $1-40/3\pi^2=-0.35$ which is unfortunately negative but an artifact of the small $N$'s chosen. Our anomalous scaling dimension is 
\be
\eta_{B^2}=1-\frac{32}{3(2N_b+N_f)\pi^2}\left(\frac{N_f}{N_b}+14\right).
\ee

\subsubsection{Superconducting order parameter}

\begin{figure}[htbp]
\centering
\begin{tikzpicture}
\node at (-1,0) {(b)};
\filldraw (0,0) circle (2pt);
\draw[thick,-<-=0.5] (0,0)--(1.2,0.75);
\draw[thick,-<-=0.5] (1.2,0.75)--(2.4,1.5);
\draw[thick,-<-=0.5] (0,0)--(1.2,-0.75);
\draw[thick,-<-=0.5] (1.2,-0.75)--(2.4,-1.5);
\draw[thick, dashdotted,->-=0.5] (1.2,-0.75)--(1.2,0.75);
\node at (0.35,0.8) {$k_1-p$};
\node at (0.,0.35) {\textcolor{blue}{$b, t$}};
\node at (0.35,-0.8) {$k_2+p$};
\node at (0.,-0.4) {\textcolor{blue}{$a, s$}};
\node at (2,1.7) {$k_1$};
\node at (1.5,1.35) {\textcolor{blue}{$b',t$}};
\node at (2.2,-1.7) {$k_2$};
\node at (1.5,-1.5) {\textcolor{blue}{$a',s$}};
\node at (1.6,0.2) {$p$};
\node at (1.6,-0.3) {$\textcolor{orange}{H_j}$};
\end{tikzpicture}
\caption{Additional contribution to the SC vertex correction, compared with fig. \ref{fig:SC_0}.}
\label{fig:SC1}
\end{figure}
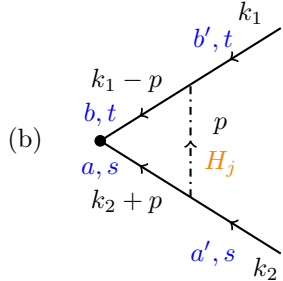

The additional diagram is figure \ref{fig:SC1}. 
The relevant integral is the same as $-I_{\lambda;2}$. Index summation gives minus three as in the gauge field correction, resulting in
\be
\begin{split}
\iota_{\text{vertex}} & =\frac{4}{2N_b \pi^2}(1+3)-\frac{8\cdot 3}{(2N_b+N_f)\pi^2} (1-\zeta)\\
& = \frac{8}{N_b \pi^2}-\frac{24}{(2N_b+N_f)\pi^2} (1-\zeta)
.
\end{split}
\ee
The dimension of this quadratic boson is thus
\be
\text{dim}\left[ B_{as} \varepsilon_{ab}  J_{st} B_{bt} \right]= (1+\eta_B)+\iota_{\text{vertex}}
=1+\frac{4}{3(2N_b+N_f)\pi^2}\left(5\frac{N_f}{N_b}-38\right),
\ee
Taking $N_f=N_b=2$, the dimension is $1-22/3\pi^2=0.26.$ 
Anomalous scaling dimension is then
\be
\iota_{B^2}=1+\frac{8}{3(2N_b+N_f)\pi^2}\left(5\frac{N_f}{N_b}-38\right).
\ee

\subsubsection{Correlation length exponent}

In addition to the diagrams presented in figures \ref{fig:self_energy}, now we also have figure \ref{fig:self_energy2} due to the $H_j$ field:
\begin{figure}[htbp]
\centering
\begin{tikzpicture}[scale=0.7]
\node at (-5,2) {(g)};
\draw[thick,->-=0.5](-4,0) -- (-2,0);
\draw[thick,->-=0.5](-2,0) -- (2,0);
\draw[thick,->-=0.5](2,0) -- (4,0);
\draw[thick,dashdotted,->-=0.5] (-2,-0.5) .. controls +(-0.4,3.5) and +(0.4,3.5) .. (2,-0.5);
\filldraw[fill=white,draw=white] (-2.5,-0.03)--(-1.5,-0.03)--(-1.5,-0.5)--(-2.5,-0.5);
\filldraw[fill=white,draw=white] (2.5,-0.03)--(1.5,-0.03)--(1.5,-0.5)--(2.5,-0.5);
\node at (-3,-0.4) {$k$};
\node at (3,-0.4) {$k$};
\node at (0,-0.4) {$k-p$};
\node at (-3,+0.5) {$\textcolor{blue}{a, s}$};
\node at (3,+0.5) {$\textcolor{blue}{a', s}$};
\node at (0,+0.5) {$\textcolor{blue}{a'', s}$};
\node at (0,1.6) {$p$};
\node at (0,2.7) {\textcolor{orange}{$H_j$}};
\end{tikzpicture}
\quad 
\quad
\begin{tikzpicture}[scale=0.7]
\node at (-5,2) {(h)};
\draw[thick,->-=0.5](-4,0) -- (0,0);
\draw[thick,->-=0.5](0,0) -- (4,0);
\draw[thick,dashed] (0,0)--(0,0.8);
\draw[thick,-<-=0.3] (0,1.6) circle (0.8);
\draw[thick, dashdotted] (-0.8,1.6) -- (0.8,1.6);
\node at (-3,-0.4) {$k$};
\node at (3,-0.4) {$k$};
\node at (-3,+0.5) {$\textcolor{blue}{a, s}$};
\node at (3,+0.5) {$\textcolor{blue}{a, s}$};
\node at (0.4,0.4) {$0$};
\node at (0,2.) {$q$};
\node at (1,1.) {$p$};
\node at (-0.4,0.4) {$\textcolor{orange}{\lambda}$};
\end{tikzpicture}
\caption{Feymann diagrams in addition to fig. \ref{fig:self_energy} that contribute to $G^{-1}_B(0)$.}
\label{fig:self_energy2}
\end{figure}
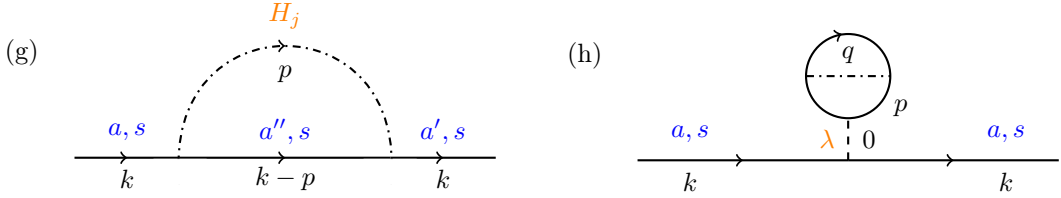
Notice that other diagrams such as replacing the dashed $\lambda$ line in diagram \ref{fig:self_energy}(d) by the dashdotted $H_z$ line will cancel each other since we need to sum over boson bubbles with different gauge indices. The self energies corresponding to the above two diagrams are 
\be
\Sigma^{(g)}= -\Sigma^{(c)},\quad 
\Sigma^{(h)}= -\Sigma^{(e)}.
\ee
Both relevant integrals have been evaluated before, leading to 
\be
\Sigma^{(g)}+\Sigma^{(h)} \rightarrow -\frac{3}{\pi^2 N_b}\lambda_g \log \lambda_g.
\ee
Combining with the $v=0$ results, we now have the modified total coefficient
\be
\alpha=-\frac{3}{N_b\pi^2}-\frac{12}{\pi^2} \left(\frac{7N_f-18N_b}{(N_f+2N_b)^2}+\frac{\zeta}{N_f+2N_b}\right).
\ee
The correlation function exponent is then (we also need to use the modified anomalous scaling dimension of $B$ in \eqref{eq:etaB2}):
\be
\nu_B\approx  1+\frac{7}{3N_b \pi^2}-\frac{20}{(2N_b+N_f)\pi^2}+\frac{12}{\pi^2}\frac{7N_f-18N_b}{(2N_b+N_f)^2}.
\ee
At $N_b=N_f=2,$ this is $\nu_B=0.037.$

\subsubsection{Boson-fermion Composite}

In the expression
$
\text{dim}[B^{\dagger}\psi]  =\text{dim}[B]+\text{dim}[\psi]+\eta_{B^{\dagger}\psi}= \frac{3}{2}+\frac{\eta_B+\eta_{\psi}}{2}+\eta_{B^{\dagger}\psi},
$
the only change compared with \eqref{eq:composite_0} lies in dim$[B]$.
The result is thus
\be
\text{dim}[B^{\dagger}\psi]  = 
\frac{3}{2} - \frac{2}{3(2N_b+N_f)\pi^2}\left(\frac{N_f}{N_b}+62\right). 
\ee
Again other gauge-invariant choices such as $\sum_{a,b} B_{as} \varepsilon_{ab} \psi_{bt}$, or with $\gamma$ matrices inserted give the same result. At $N_b=N_f=2$, we have dim$[B^{\dagger}\psi]=\frac{3}{2}-\frac{7}{\pi^2}=0.79.$

\section{Useful integrals}
\label{app:detail}
Below we present some details of the integrals that appear in the main text. 

We first present more details for the derivation of the effective action in section \ref{subsec:v=0}. Expansion of the matrix log gives, in the bosonic sector, 
\be
\begin{split}
\tr \ln \mathcal{G}^{-1}_b G_B=  & 2N_b \int \frac{d^3p}{(2\pi)^3} \frac{d^3q}{(2\pi)^3}  \Bigg\{  G(q) \delta_{p0}  \int \frac{d^3 p'}{(2\pi)^3} A_{\alpha\mu}(p')A_{\alpha}^{\mu}(-p')\\
& \quad \quad -\frac{1}{2} G(q)G(q-p) \bigg[
- \lambda(p)\lambda(-p)+  \sum_j (2q-p)_{\mu} A_{\alpha}^{\mu}(p)  (2q-p)_{\nu} A_{\alpha}^{\nu}(-p) \bigg]\Bigg\}.\\
\end{split}
\ee
For the fermion sector we have
\be
\tr\ln \mathcal{G}^{-1}_f G_{\psi} = -\frac{N_f}{2}  \int \frac{d^3p}{(2\pi)^3} \frac{d^3q}{(2\pi)^3}
\tr[G_{\psi}(q) \gamma^{\mu} A^{\alpha}_{\mu}(p) \sigma_{\alpha} G_{\psi}(p+q)\gamma^{\nu}A^{\beta}_{\nu}(-p)\sigma_{\beta}].
\ee
The integrals can be evaluated, we summarize the results here:
\be
\begin{split}
& \int \frac{d^3q}{(2\pi)^3} \frac{1}{q^2+ \bar{\lambda}}  =-\frac{\sqrt{\bar{\lambda}}}{4\pi},\quad \quad \int \frac{d^3q}{(2\pi)^3} \frac{\tr[\gamma^{\mu}\slashed{q}\gamma^{\nu}(\slashed{p}+\slashed{q})]}{q^2(p+q)^2}  =-\frac{1}{16q},\\
& \int \frac{d^3q}{(2\pi)^3} \frac{1}{(q^2+\bar{\lambda})\big( (q-p)^2+\bar{\lambda}\big)}  =\frac{1}{4\pi p}\arctan \frac{p}{2\sqrt{\bar{\lambda}}},\\
& \int \frac{d^3q}{(2\pi)^3} \frac{(2q-p)_{\mu}(2q-p)_{\nu}}{(q^2+\bar{\lambda})\big( (q-p)^2+\bar{\lambda}\big)}  = -\left(\delta_{\mu\nu}+\frac{p_{\mu}p_{\nu}}{p^2}\right) \frac{\sqrt{\bar{\lambda}}}{4\pi}  -\bigg(\delta_{\mu\nu}-\frac{p_{\mu}p_{\nu}}{p^2}\bigg)\bigg(\frac{4\bar{\lambda}+p^2}{8p\pi} \arctan \frac{p}{2\sqrt{\bar{\lambda}}}\bigg).
\end{split}
\label{eq:integrals}
\ee
The leading correction to the free energy is thus \eqref{eq:eff_F0}. Notice the first and second order bosonic contributions to $\Pi_A$ combine to give a simple expression.

Next we evaluate the integrals $I_{A;i}$ and $I_{\lambda;i}$ with $i=1,2$ that appear in the main text. 
\be
\begin{split}
I_{A;1}= & \int \frac{d^3p}{8\pi^3} \left[ G_B(k+p)D_{\mu\nu}(-p)(2k+p)_{\mu}(2k+p)_{\nu}\right] \mid_{\bar{\lambda}=0}\\
=\ &  -\frac{16}{2N_b+N_f}  \int \frac{d^3p}{8\pi^3}\ \frac{4k_{\mu}k_{\nu}+2k_{\mu} p_{\nu}+2p_{\mu} k_{\nu}+p_{\mu} p_{\nu}}{p(k+p)^2}\left(\delta_{\mu\nu}-\zeta \frac{p_{\mu}p_{\nu}}{p^2}\right)\\
=\ &  -\frac{16}{2N_b+N_f}  \int \frac{d^3p}{8\pi^3} \frac{1}{p(k+p)^2} \Big[
(4k^2+4k\cdot p+p^2)-\frac{\zeta}{p^2} \big(4 (k\cdot p)^2+4p^2(k\cdot p)+p^4\big)
\Big]\\
=\ &  -\frac{16}{2N_b+N_f} \int \frac{dp}{4\pi^2}\int d\theta  \frac{p \sin\theta }{(k^2+p^2+2kp\cos\theta)} \Big[
(4k^2+4kp\cos\theta +p^2)-\zeta \big(4 k^2\cos\theta^2+4kp\cos\theta+p^2\big)
\Big]\\
=\ & -\frac{16}{2N_b+N_f} \int \frac{dp}{4\pi^2} \Big[ 
4p+\frac{2k^2-p^2}{2k}\log\left(\frac{k+p}{k-p}\right)^2
+2\zeta \frac{k^2-p^2}{p}- \zeta \frac{k^3}{2p^2}\log\left(\frac{k+p}{k-p}\right)^2
\Big]\\
\rightarrow \ & -\frac{4}{(2N_b+N_f)\pi^2} \left(\frac{10}{3}+2\zeta \right)k^2\log k, 
\end{split}
\ee
where the right arrow in the last line means we are extracting the $k^2\log k$ divergence. 

For \eqref{eq:I_lambda_1}, we have
\be
\begin{split}
I_{\lambda;1} = & \i^2 \int \frac{d^3p}{8\pi^3} \left[G_B(k+p)D_{\lambda}(-p)\right]\big|_{\bar{\lambda}=0}\\
= &\ \frac{8}{2N_b} \int \frac{d^3p}{8\pi^3} \frac{p}{(k+p)^2}\\
= &\ \frac{8}{2N_b} \int \frac{dp}{4\pi^2}\int \sin\theta d\theta \frac{p^3}{k^2+p^2+2kp\cos\theta}\\
= &\  \frac{2}{2N_b \pi^2}\int dp \frac{p^3}{2kp}\log\left(\frac{k+p}{k-p}\right)^2\\
\rightarrow &\ \frac{2}{3 N_b\pi^2}k^2\log k.
\end{split}
\ee

As for \eqref{eq:IA2}, we have
\be
\begin{split}
I_{A;2} \rightarrow &\ \frac{16}{2N_b+N_f} \int \frac{d^3p}{8\pi^3} \frac{(2k-p)_{\mu}(2k-p)_{\nu}}{(k-p)^4} \frac{1}{p} \left(\delta_{\mu\nu}-\zeta \frac{p_{\mu}p_{\nu}}{p^2}\right)\\
=\ & \frac{16}{2N_b+N_f} \int \frac{d^3p}{8\pi^3}\frac{1}{p} \left[ \frac{(2k-p)^2}{(k-p)^4}-\zeta \frac{[(2k-p)\cdot p]^2}{p^2 (k-p)^4}\right] \\
=\ & \frac{16}{2N_b+N_f} \int \frac{dp}{4\pi^2}\int d\theta \sin\theta \left[
p\frac{4k^2-4kp\cos\theta+p^2}{(k^2+p^2-2kp\cos\theta )^2}-\zeta \frac{(2kp\cos\theta-p^2)^2}{p(k^2+p^2-2kp\cos\theta)^2}
\right]\\
\rightarrow \ & \frac{16}{2N_b+N_f} \frac{1}{4\pi^2} (1-\zeta) (-2\log k) = - \frac{8}{(2N_b+N_f)\pi^2}(1-\zeta)\log k. 
\end{split}
\ee
In the last line we have again extracted the term proportional to $\log k$.

In \eqref{eq:PH_lambda_0}, 
\be
\begin{split}
I_{\lambda;2} = & \i^2\int \frac{d^3p}{8\pi^3} G_B(k_1-p)G_B(-k_2+p)D_{\lambda}(p)\\
= & - \frac{8}{2N_b} \int \frac{d^3p} {8\pi^3} \frac{p}{(k_1-p)^2}\frac{1}{(k_2-p)^2}.\\
= & - \frac{8}{2N_b} \int \frac{dp}{4\pi^2}\int d\theta \frac{p^3 \sin\theta }{(k^2-2kp\cos\theta+p^2)^2} \\
= & -\frac{8}{2N_b}\int \frac{dp}{4\pi^2} \frac{2p^3}{(k^2-p^2)^2}\\
\rightarrow & -\frac{8}{2N_b}\frac{1}{4\pi^2}(-2\log k)= +\frac{2}{N_b\pi^2}\log k.
\end{split}
\ee

\bibliography{refs}

\end{document}